\documentclass[10pt,jurnal]{IEEEtran}
\usepackage{balance}
\usepackage{amssymb}
\usepackage{stmaryrd}
\usepackage{cite}
\usepackage{color}
\usepackage{multirow}
\usepackage{graphicx,times}
\usepackage{indentfirst}
\usepackage{CJK}
\usepackage{amsmath}
\usepackage{amsfonts}
\usepackage{txfonts}
\usepackage{mathrsfs}
\usepackage{algorithm}  
\usepackage{algorithmic}  
\usepackage{bm}
\usepackage{subfigure}
\usepackage{theorem}

\newtheorem{theorem}{Theorem}
\newtheorem{definition}{Definition}

\usepackage{color}

\begin{document}
\title{A Two-stage Game Framework to Secure Transmission in Two-Tier UAV Networks}
\author{
	\IEEEauthorblockN{Mengnian Xu, Yanjiao Chen,~\IEEEmembership{Member,~IEEE}, Wei Wang,~\IEEEmembership{Senior Member,~IEEE}}
	\thanks{This work was supported in part by the National Natural Science Foundation of China under Grants 62071194, 91738202, 61972296 and 61702380, Young Elite Scientists Sponsorship Program by CAST under Grant 2018QNRC001, National Key R\&D Program of China under Grant 2017YFE0121500, and Wuhan Advanced Application Project under Grant 2019010701011419.}
	\thanks{M. Xu and W. Wang are with the School of Electronic Information and Communications, Huazhong University of Science and Technology, Wuhan 430074, China (e-mail:\{mengnian, weiwangw\}@hust.edu.cn).}
	\thanks{Y. Chen is with the School of Computer Science, Wuhan University, Wuhan 430072, China (e-mail: chenyj.thu@gmail.com).}
	
}

\maketitle
\begin{abstract}
The multi-UAV network is promising to extend conventional networks by providing broader coverage and better reliability. Nevertheless, the broadcast nature of wireless signals and the broader coverage expose multi-UAV communications to the threats of passive eavesdroppers.
Recent studies mainly focus on securing a single legitimate link, or communications between a UAV and multiple ground users in one/two-UAV-aided networks, while the physical layer secrecy analysis for hierarchical multi-UAV networks is underexplored.
In this paper, we investigate a general two-tier UAV network consisting of multiple UAV transmitters (UTs) and  multiple UAV receivers (URs) in the presence of multiple UAV eavesdroppers (UEs). 
To protect all legitimate UT-UR links against UEs at the physical layer, we design a two-stage framework consisting of a UT-UR association stage and a cooperative transmission stage. Specifically, we formulate the secure transmission problem into a many-to-one matching game followed by an overlapping coalition formation (OCF) game, taking into account the limited  capabilities and the throughput requirements of URs, as well as the transmission power constraints of UTs. A matching algorithm and an OCF algorithm are proposed to solve these two sequential games whose convergences and stabilities are guaranteed theoretically. Simulation results show the superiority of our algorithms and the effectiveness of our two-stage game framework in the terms of secrecy performance.     
\end{abstract}

\section{Introduction}\label{sec:intro}
Unmanned aerial vehicles (UAVs) have been rapidly developed over the last decades based on their widely known advantages such as low cost and on-demand deployment. Since a multi-UAV network can cooperatively complete the mission more efficiently with larger coverage and resilience to node failure, the use of multi-UAV networks to perform various sensing tasks has recently drawn much attention, such as precision agriculture, city traffic monitoring, and disaster management~\cite{doering2014design,elloumi2018monitoring,dey2017ad}. 
In a multi-tier UAV architecture~\cite{sekander2018multi}, different types of UAVs that have unique features such as maximum flight altitude, communication coverage, computing ability and durability, can take different roles, which enables more functional diversity for next-generation wireless communications.

Although the broadcast nature of wireless signals and the broad coverage provide efficient collaboration in a UAV network, they also expose UAV-to-UAV communications to malicious eavesdroppers, especially the more flexible UAV eavesdroppers. A multi-UAV network usually collects vital or confidential messages, such as the information for monitoring an airport or a nuclear station, which may be intercepted by criminals or terrorists to commit crimes. Therefore it is increasingly urgent and necessary to ensure secure communications against eavesdroppers in multi-UAV networks.    
 
In contrast to traditional cryptographic-based methods to secure transmission, which are computationally costly and unsuitable for resource-constrained UAVs, physical layer security (PLS) has been an effective alternative. Nevertheless, although PLS in wireless communications has received much attention, there have been few studies on protecting multi-UAV communications against eavesdropping. Most existing PLS schemes in UAV-aided networks focus on securing a single legitimate source-destination communication link, leaving the secure transmission issue in hierarchical multi-UAV communications underexplored. Specifically, they either protect a single UAV-ground link from terrestrial eavesdropping~\cite{cui2018robust,zhang2019securing}, or secure a single traditional terrestrial link against ground eavesdroppers~\cite{zhou2018improving} or UAV eavesdroppers~\cite{tang2019secrecy}, in which only one UAV acts as a mobile BS/relay/jammer to improve PLS. Furthermore, secure communications between a UAV-BS and multiple ground users in the presence of ground eavesdroppers have been studied in~\cite{zhou2019uav}, where an additional UAV-jammer is used to disturb the eavesdroppers, and PLS is improved by jointly optimizing the trajectories and transmit power of UAV-BS and UAV-jammer. However, these two UAVs are assumed to fly at a fixed altitude, which simplifies the analysis but limits the applicability in practice. In addition, it is difficult to apply UAV trajectory and transmit power control method directly into multi-UAV networks since then the collision avoidance becomes non-negligible and more complex.   

In this paper, we take the first step to investigate a two-tier UAV network consisting of multiple UAV transmitters (UTs) and multiple UAV receivers (URs), where each UT collects sensing data from its coverage of interest and then delivers the data to a UR for further processing. In this case, multiple UAV eavesdroppers (UEs) intend to wiretap on the legitimate links. Our goal is to secure all legitimate UT-UR links against UEs in this network by jointly considering the UT-UR association and the cooperative transmission scheduling. On one hand, by properly associating UTs and URs, short-distance LoS links for data transmission can be proactively established which benefit both secrecy performance and the UR throughput requirements. On the other hand, by performing cooperative beamforming in which case a UT-source and some other UT-relays transmit together towards the intended UR~\cite{dong2010improving}, the achievable secrecy rate can be significantly improved. Thereby we design a two-stage framework consisting of a UT-UR association stage and a cooperative transmission stage. To achieve this two-stage framework, we need to tackle the following challenges. 
    
In the first UT-UR association stage, the limited processing capability of each UR should be considered, thus the number of UTs that a UR can serve is limited. Moreover, the throughput requirements of URs should also be addressed in the association process, which may conflict with the need for high network secrecy performance.
The second cooperative transmission stage aims to find an effective relay selection strategy for each UT under a transmission power constraint, to maximize the total secrecy utility of the whole network. 
However, such a UT-UR association problem and a relay selection problem are non-trivial to solve because both of them are NP-hard problems.

To tackle the above challenges, we formulate the UT-UR association problem as a many-to-one matching game, in which UTs and URs make their matching decisions based on their individual preferences. Then, to model the complex cooperative behaviors among the UTs in the cooperative transmission stage, we formulate the relay selection problem as an overlapping coalition formation (OCF) game, in which a UT can join multiple coalitions to assist multiple UTs' transmission in different time slots. In this way, by solving these two sequential games, the network can achieve a stable association structure and a stable overlapping coalition structure to perform secure cooperative transmission. 
In our proposed multi-UAV network, UTs and URs are viewed as selfish and rational agents aiming to improve their own utilities through interaction and cooperation. This scenario fits the future heterogeneous UAV network paradigm where different companies with potential conflict interests launch their UAVs. As game players, UTs and URs make their own decisions in a distributed manner, which provides the network with self-organizing capability and adaptability to diverse circumstances, thus benefiting the management and control of multi-UAV networks.

A comprehensive survey of the application of game theory to UAV networks is presented in~\cite{mkiramweni2019survey}. Compared with centralized optimization methods that usually require frequent information exchange between UAVs and the central controller, game-based methods enable individual UAVs to make their own decisions, thus reducing the communication overhead.
 
Our main contributions can be summarized as follows.

\begin{itemize}
	\item We design a two-stage framework to protect all legitimate communication links against UAV eavesdroppers in a hierarchical multi-UAV network from a physical layer perspective.
	\item We formulate the UT-UR association problem into a many-to-one matching game. Based on the characteristics of our matching problem, we design a matching algorithm to achieve a pairwise stable matching result with higher social welfare of UTs and URs.
	\item We formulate the relay selection problem as an OCF game. We also propose an OCF algorithm that maximizes the collaboration between UTs and fully seeks the optimal structure, to achieve a stable coalition structure for performing cooperative transmission.
	\item Extensive simulations under various system parameters prove the superiority of our algorithms, and the effectiveness of our two-stage game framework in the terms of secrecy performance. 
\end{itemize}


\section{Related Work}\label{sec:related work}
In this section, we show that our technical work differs from some existing works related to matching theory and coalition formation game theory in recent years, and we summarize the differences as follows.

\textbf{Matching Theory.} Our matching algorithm differs from the existing works. First, we have distinctly different flow design. Quite a few works~\cite{leanh2017matching,hamidouche2017popular,pham2017traffic} on matching theory are extensions of the classical deferred acceptance (DA) algorithm, whose convergence and stability are easily guaranteed. In comparison, our proposed algorithm consists of two phases, i.e., a preliminary interaction and a swapping operation. Note that the DA algorithm is used as a baseline for comparison with our algorithm in our paper. Second, we have distinct matching rules. The proposed algorithms in~\cite{semiari2015matching,el2016matching,kazmi2017mode,kazmi2020distributed} differ from the conventional DA algorithm. However, none of these algorithms are applicable to our problem since our preference functions have different characteristics. The matching rules in~\cite{semiari2015matching,el2016matching,kazmi2017mode,kazmi2020distributed} include that one cannot propose to the same object twice and one can remove current matched objects. Different from these rules, our matching rules are specifically designed for our particular preference functions, aiming to yield higher social welfare. 

\textbf{Coalition Formation Game Theory.} Our coalition formation algorithm differs from the existing works. Some works~\cite{niyato2011controlled,saad2011coalitional} focused on the non-overlapping coalition formation game model, in which the players can only form disjoint coalitions. In our paper, we model the relay selection problem as an OCF game that allows a player to participate in multiple coalitions to cooperate with more UTs, hence improving the performance gain via a more complex coalition structure. Apart from the differences in game formulation, our algorithm design is also different. For example, the solution in~\cite{niyato2011controlled} is based on split-and-merge strategies, which are only suitable for the disjoint coalition formation games. And the algorithm in~\cite{saad2011coalitional} is only based on a switching rule, which greatly limits the search for the potential optimal coalition structure. In comparison, our proposed algorithm can reach a stable overlapping coalition formation. It is worth mentioning that we have used the disjoint coalition game model and the commonly-used merge-and-split strategies as one of our baseline schemes for comparison.

Our algorithm design is also different from the related works on OCF games~\cite{zhang2014coalitional,wang2013overlapping}. First, we have different initial states, which are important for the evolution of coalition formation. In~\cite{zhang2014coalitional,wang2013overlapping}, the initial overlapping coalition structure (OCS) is given as a set of singleton coalition. Unlike the simple singleton coalitions, we identify all potential partners of a UT and put them into a coalition. Besides, we add a corrective action to obtain a preliminary structure, which takes into account the constraints of the limited communication range and the necessary number of coalition members. In this way, our initialization can maximize collaboration between players. Second, we have different operation rules for each player to update OCS. Specifically, in~\cite{zhang2014coalitional}, only the joining operation and the singleton coalition formation operation are allowed, and in~\cite{wang2013overlapping}, only the switching operation and the singleton coalition formation operation are allowed, which may hinder the formation of desirable coalition structures in the iteration process. In comparison with the limited operation, we provide each player with more options to search for a better structure.

\begin{figure}[t]
	\centering
	\includegraphics[width=6.5cm,height=4.4cm]{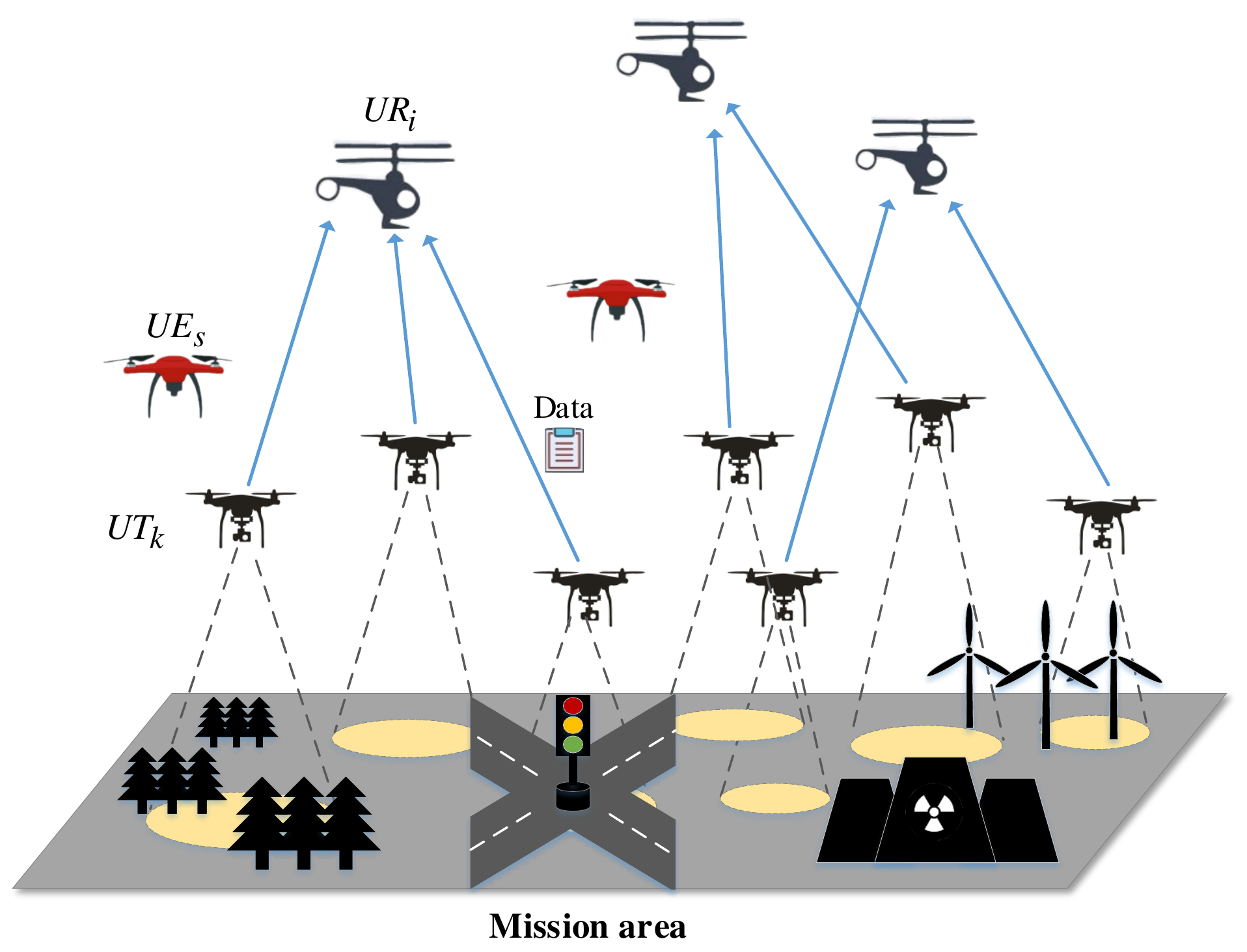}\vspace{-0.3cm}
	\caption{A two-tier UAV network consisting of multiple UTs and URs in the presence of multiple UEs. The UTs and URs are deployed to execute various sensing tasks, e.g. precision agriculture, city traffic monitoring.} \vspace{-0.4cm}
	\label{fig:network}
\end{figure}

\section{System Model}\label{sec:system model}
\subsection{Scenario Description}
We investigate a two-tier UAV network consisting of $M$ URs, $N$ UTs, and $S$ UEs in the air, which are defined as $\mathcal{M}\triangleq\{1,2,...,M\}$, $\mathcal{N}\triangleq\{1,2,...,N\}$ , and $\mathcal{S}\triangleq\{1,2,...,S\}$ respectively. We denote by $UT_k$ the UT $k$, by $UR_i$ the UR $i$, and by $UE_s$ the UE $s$ in this paper. As illustrate in Fig.~\ref{fig:network}, there are $N$ UTs collecting data from their areas of coverage, and $M$ URs flying at a higher altitude to receive and process data from the UTs. A UR can serve multiple UTs while a UT can only choose one UR as its receiver. We assume that this network adopts a time division multiple access (TDMA) protocol which is a typical transmission protocol, and our proposed algorithm can be readily extended to the case of other protocols such as frequency division multiple access (FDMA). Each $UT_k$ is allocated with a fixed time slot $TS_k$ for its data transmission, $\forall k=1,2,\cdots,N$. Without loss of generality, we consider that all UTs and URs operate over a common frequency band $W$. Thus, when this network is well timed and synchronized, the transmission is collision-free, since there is only one UT transmitting data in each time slot and each UR will only receive data from its matched UTs.

All the UTs, URs and UEs are assumed to work in a half-duplex mode, be equipped with a single omni-directional antenna. Let $h_{ki}$, $h_{ks}$ represent the channel gain of UT-UR link and UT-UE link respectively, where $k \in \{1,2,...,N\}$, $i \in \{1,2,...,M\}$, and $s \in \{1,2,...,S\}$. We also assume that the global location information, including the UEs, is completely known as the locations of eavesdroppers can be estimated via an optical camera or synthetic aperture radar (SAR) equipped on the UAV~\cite{zhou2019uav,zhang2019securing}. UTs can further estimate the channel gain based on the obtained location information since the LoS channel gains only depend on the distance. Unlike the air-to-ground propagation model which needs to calculate both LoS and NLoS pathloss, UAV-to-UAV communications are indeed air-to-air propagations, thus there is no need to consider NLoS channel due to few obstacles in the air and the dominance of LoS link~\cite{goddemeier2015investigation}. For simplicity, we model the LoS channel as $h_{mn}=d_{mn}^{- \alpha /2} e^{j \theta}, mn \in \{ki,ks\}, \forall k, i, s$, where $d_{ki}$ and $d_{ks}$ are the distances from $UT_k$ to $UR_i$ and $UE_s$ respectively. $\alpha$ is the path loss exponent, and $\theta$ is a random phase distributed within $\left[0,2 \pi\right)$. 

Note that we focus on a static scenario where UAVs stay static or quasi-static performing various sensing tasks. The dramatical fluctuation of channels caused by the dynamic flight states of the UAV~\cite{xiao2019sensor,he2019state} is beyond the scope of our study. The impact of mobility and flight states of the UAV on physical layer security is an interesting research topic and it may be our future work.

\subsection{Cooperative Data Transmission}
We consider a common cooperative relaying protocol referred to as decode-and-forward (DF). The complete data transmission can be divided into two phases. In the first phase, the UT source broadcasts its message to some UT relays, which is referred to as the broadcast phase. In the second phase, all the relays together with the UT source cooperatively transmit a weighted version of the re-encoded message to the intended UR receiver, which is referred to as the transmission phase (note that the intended UR receiver refers to the desired UR receiver of the UT source). For each $UT_k$'s message transmission, $TS_k$ is evenly divided into two sub-slots for the broadcast phase and the transmission phase, respectively. 

We consider there is an overall power budget $P_{0}$ for each $UT_k$'s transmission, which constrains $UT_k$ and all its UT relays. We denote $P_{b}$ as the transmit power in the broadcast phase, and $P_{t}$ as the total transmit power of the source and all the relays in the transmission phase. Obviously, we have $0 \leq P_b, P_t \leq P_0$. Suppose there are $n-1$ relays assisting $UT_k$ to transmit towards $UR_i$, the beamforming weights are denoted by a $n\times1$ vector $\boldsymbol{w}$ ($n-1$ relays plus the source). In addition, the channel gain vector from the $n$ UTs to $UR_i$ and $UE_s$ are denoted by $\boldsymbol{h}_{TR_i}=\left[h_{T_{1}R_i},h_{T_{2}R_i},\cdots,h_{T_{n}R_i} \right]^{\top}$, $\boldsymbol{h}_{TE_s}=\left[h_{T_1E_s},h_{T_2E_s},\cdots,h_{T_nE_s} \right]^{\top}$, respectively, where $(\cdot)^{\top}$ means transpose. We define a $n\times S$ channel matrix $\boldsymbol{H}_{TE}$, which represents the channels between $n$ UTs and $S$ UEs. Thus, when $UT_k$ together with its relays transmit a weighted version of its message, the received signal at $UR_i$ and $UE_s$ are given by 
\begin{align}
{y_{R_i}}=& \boldsymbol{h}_{TR_i}^{\dagger} \boldsymbol{w} \hat{x} + n_{R_i} \\
{y_{E_s}}=& \boldsymbol{h}_{TE_{s}}^{\dagger} \boldsymbol{w} \hat{x} + n_{E_{s}}, 
\end{align}
where $(\cdot)^{\dagger}$ represents conjugate transpose, $\hat{x}$ is the re-encoded symbol which is normalized, i.e., $\mathbb{E}\{\left| \hat{x}\right| ^{2}\}=1$, and $n_{R_{i}}$ and $n_{E_{s}}$ represent white complex Gaussian noise with zero-mean and variance $\sigma^{2}$ at the $UR_i$ and $UE_s$, respectively. 
 
Then in the presence of multiple UEs, the achievable secrecy rate in the transmission phase can be expressed as
\begin{align}
{C_{k}}=& \left[\log(1+\gamma_{R_i})-\max_{s\in\mathcal{S}}\log(1+\gamma_{E_s})\right] ^{+} \notag \\
=& \left[\log(1+\frac{\left|\boldsymbol{w}^{\dagger} \boldsymbol{h}_{TR_i}\right|^{2} }{\sigma_i^{2}})-\max_{s\in\mathcal{S}}\log(1+\frac{\left|\boldsymbol{w}^{\dagger} \boldsymbol{h}_{TE_{s}}\right|^{2} }{\sigma_s^{2}})\right]^{+}  \label{equation6},
\end{align}
where $\left[ a\right] ^{+}$ represents $\max(a,0)$, and $\gamma_{R_i}$ and $\gamma_{E_s}$ represent the signal-to-noise-ratio (SNR) at $UR_i$ and $UE_s$, respectively. For simplicity, we consider an extreme case in which we completely null out signals at all UEs, i.e., $\boldsymbol{w}^{\dagger} \boldsymbol{H}_{TE}=\boldsymbol{0}_{1\times S}$, which is referred to as \textit{null-steering beamforming}~\cite{dong2010improving}. Then, the second item in~(\ref{equation6}) is zero. We aim to maximize the achievable secrecy rate with the constraints of transmit power budget and nulling out signals at UEs. To get the optimal weight vector, we formulate this problem as
\begin{align}
&\boldsymbol{w}^{*}=\arg\max_{\boldsymbol{w}} \left|\boldsymbol{w}^{\dagger} \boldsymbol{h}_{TR_i}\right|^{2}  \notag \\
&s.t.\quad
\begin{cases}
\boldsymbol{w}^{\dagger} \boldsymbol{w} \leq P_{0}\\
\boldsymbol{w}^{\dagger} \boldsymbol{H}_{TE}=\boldsymbol{0}_{1\times S}.
\end{cases}   \label{equation7}
\end{align}
There is a closed-form solution for the above optimization problem, which is given by~\cite{dong2010improving} 
\begin{equation}
\boldsymbol{w}^{*}=\frac{\sqrt{P_{0}}}{\left\| \left(\boldsymbol{I}_{n}-\boldsymbol{U}_{TE}\right) \boldsymbol{h}_{TR_i} \right\| } \left(\boldsymbol{I}_{n}-\boldsymbol{U}_{TE}\right) \boldsymbol{h}_{TR_i},
\end{equation} 
where $\boldsymbol{I}_{n}$ is the $n\times n$ identity matrix, $\left\|\boldsymbol{a} \right\| $ is the 2-norm of vector $\boldsymbol{a}$, and $\boldsymbol{U}_{TE}\triangleq \boldsymbol{H}_{TE} \left( \boldsymbol{H}_{TE}^{\dagger} \boldsymbol{H}_{TE}\right)^{-1} \boldsymbol{H}_{TE}^{\dagger}$. Note that in order to successfully null the signal at all $S$ UEs and obtain the optimal weight vector, we need $n>S$ here, which means that a UT needs at least additional $S$ UT relays to execute cooperative beamforming together. 

\section{A Many-to-one Matching Game for UT-UR Association}\label{sec:UT-UR Association}
\subsection{Many-to-one Matching Game Formulation}
In the first stage, we need to optimally match multiple UTs (no more than the UR's quota, i.e., the maximum number of UTs the UR can serve) with one UR considering their different characteristics and requirements. Specifically, for each UT, we view its secrecy rate performance as its benefit, and for each UR, we regard its average throughput as its benefit. We aim to maximize the social welfare of both URs and UTs in a self-organized manner. Before showing the proposed matching algorithm, we introduce several basic definitions~\cite{gu2015matching}.
     
\begin{definition}
In our scenario, a many-to-one matching is a function $\Phi$: $\mathcal{M}\cup \mathcal{N} \to \mathcal{M}\cup \mathcal{N}$, such that
\end{definition}
	
\begin{enumerate}
	\item $\Phi\left( UR_{i}\right) \subseteq \mathcal{N} $, and $\left| \Phi(UR_{i}) \right| \leq Q_i , \forall i\in \mathcal{M}$
	\item $\Phi\left( UT_{k}\right) \in \mathcal{M} $, and $\left| \Phi(UT_{k}) \right| =1 , \forall k\in \mathcal{N}$
	\item $\Phi\left( UR_{i}\right)=UT_{k} \Leftrightarrow \Phi\left( UT_{k}\right)=UR_{i} ,\forall i\in \mathcal{M},\forall k\in \mathcal{N},$
\end{enumerate}
where $Q_i$ represents the quota of $UR_i$. These three conditions imply that each UR can be matched with multiple UTs while each UT can match only one UR.

\begin{definition}
Given two disjoint and finite sets of players, $\Theta=\{\theta_{p}\}_{p=1}^{|\Theta|}$ and $\Omega=\{\omega_{q}\}_{q=1}^{|\Omega|}$, a preference relation $\succ$ is a complete and transitive binary relation between these two sets. The expression $\omega_{q} \succ_{\theta_{p}} \omega_{q^{'}}$ imply that player $\theta_{p}$ prefer $\omega_{q}$ over $\omega_{q^{'}}$, similarly, $\theta_{p} \succ_{\omega_{q}} \theta_{p^{'}} $ imply player $\omega_{q}$ prefer $\theta_{p}$ to $\theta_{p^{'}}$. 
\end{definition}

To quantify the degree of preference, we employ preference functions. As mentioned before, each UT aims to secure the communication between itself and a UR as much as possible, thus it prefers the UR who could bring it better secrecy performance. Hence, we design the preference of $UT_k$ over $UR_i$ as the secrecy rate under direct transmission
\begin{align}
U_{k}^{i}= \left[\log(1+\frac{P_{0} \left|h_{ki}\right|^{2} }{\sigma_i^{2}})-\max_{s\in\mathcal{S}}\log(1+\frac{P_{0} \left|h_{ks}\right|^{2} }{\sigma_s^{2}})\right]^{+}  \label{equation9},
\end{align}
and if there is $U_{k}^{i} > U_{k}^{i^{'}}$, which is equal to $UR_i \succ_{UT_k} UR_{i^{'}}$, then $UT_k$ prefers $UR_i$ to $UR_{i^{'}}$ in the matching process. 

Next, we design the preference function of $UR_i$. We view the average throughput of a UR as the average receiving rate within the time for receiving data. The set of UTs matched with $UR_i$ is denoted by $T_{i}$. Then the preference of $UR_i$ over the set $T_i$ is given as
\begin{equation}
U_{i}^{T_{i}}= \frac{W}{|T_{i}|} \sum_{j \in T_{i}} \log(1+\gamma_{ji}),\,\, T_{i} \subseteq \mathcal{N},\,\, |T_{i}|\leq Q_i,
\end{equation}
where $W$ is the shared bandwidth, $\gamma_{ji}$ represents the SNR of $UT_j$-$UR_i$ link. This preference function implies that, during the matching process, each UR prefers those UTs who could bring higher SNR under the constraint of quota $Q_i$. 

To better define an important definition \textit{a stable matching}, we first explain the notion of a blocking individual and a blocking pair. A blocking individual means that there exists a player who prefers to be unmatched over matching the current player under $\Phi$, in which case we say this matching is blocked by an individual. A blocking pair $(\theta_{p},\omega_{q})$ means that both $\theta_{p}$ and $\omega_{q}$ can get higher utility if they match with each other, compared to their current match, in which case we say this matching is blocked by a pair $(\theta_{p},\omega_{q})$. The conventional stability of a matching can be defined as stable if it is not blocked by any individual or pair.

\subsection{Solution for the Matching Game}
Based on the preferences design, we can see that each UT's preference is fixed and only dependent on the matched UR. However, the profit a UR can obtain is dependent on the set of matched UTs. The preferences of each UR is variable as the matching structure changes in each iteration and accepting more UTs does not necessarily bring greater benefits to a UR, which makes the classical DA algorithm unsuitable and this matching problem becomes challenging. 

Obviously, for a rational UR in each iteration, it intends to accept only the most preferred UT among the combined pool of old partners and new applicants, rejecting all the rest. However, this will lead to $N-M$ UTs unmatched eventually, which is not allowed in our work. To avoid this and achieve higher social welfare, we set up a few constraints in the matching process. The first one is that each UR can't kick out the existing matched partners. Second, a UT has a second chance to propose to the UR who has rejected it before. Three, each UR should accept the UTs applying to it for the second time or more as much as possible, of course, under a condition of not exceeding its quota.

Before delving into our proposed algorithm, we first re-define a new preference of $UR_i$ over $UT_k$ as $\tilde{U}_{i}^{k}= W\log(1+\gamma_{ji}),\,\,i \in \mathcal{N}$, which is only dependent on the specific $UT_k$, nothing to do with other UTs. We refer to it as the reference preference (RP) of $UR_i$ over $UT_k$ in the following. Moreover, we refer to the open positions to accommodate new applicants as \textit{seat}, and denote the current capacity for new UTs of $UR_i$ by $e_i$.

\begin{algorithm}[t] \footnotesize
	\caption{UT-UR Many-to-one Matching Algorithm.}
	\label{alg:matching} 
	\begin{algorithmic}
		\STATE \textbf{Data:} $Q_i$, $h_{ki}$, $h_{ks}$, $P_0$, $W$, $\sigma$\\
		\STATE \textbf{Result:} $\Phi$  
		\STATE 1.\textbf{Initialization:} $\mathcal{T}_{unmatch}=\mathcal{N}$; $e_i=Q_i$, $\forall i\in \mathcal{M}$
		\STATE Calculate preference lists:
		$\mathcal{PT}_k=\{U_{k}^{1}, U_{k}^{2}, \cdots, U_{k}^{M}\}, \forall k\in \mathcal{N}$;
		\STATE Calculate RP lists:
		$\mathcal{PR}_i=\{\tilde{U}_{i}^{1}, \tilde{U}_{i}^{2}, \cdots, \tilde{U}_{i}^{N}\}, \forall i\in \mathcal{M}$;
		\STATE 2.\textbf{Phase I:} Obtain a preliminary matching $\tilde{\Phi}$.
		\REPEAT
		\FOR{all $UT_k \in \mathcal{T}_{unmatch}$} 
		\STATE propose to the current most preferred UR based on $\mathcal{PT}_k$.
		\ENDFOR
		\FOR{$\forall UR_i$ that receive proposal} 
		\IF{$e_i=0$} 
		\STATE reject all applicants.
		\ELSE       
		\STATE divide the applicants into two categories: first-time applicants ${A}_i$, and second-time (or more) applicants ${B}_i$. If {$|{B}_i|>=e_i$}, $UR_i$ accept $e_i$ preferred applicants in ${B}_i$, and reject others. If {$|{B}_i|< e_i$}, $UR_i$ accept all members in ${B}_i$ and some applicants in ${A}_i $ according to $\mathcal{A}_i$, which is determined by Algorithm~\ref{alg:select}, and reject others.                  
		\ENDIF
		\ENDFOR
		\STATE \textbf{Update $\Phi$, $e_i$ and $\mathcal{T}_{unmatch}$} 
		\STATE \textbf{Record rejection:} for each rejected $UT_k$, record the number of times $UR_i$ rejects it and update $\mathcal{PT}_k: U_k^i=U_k^i-\delta$.
		\UNTIL{$\mathcal{T}_{unmatch}=\emptyset$}
		\STATE 3.\textbf{Phase II:} Swapping-matching operation. $\Phi \leftarrow \tilde{\Phi}$
		\REPEAT 
		\STATE Search for approved swapping pairs:$\{UT_s,UT_t,UR_h,UR_g\}$
		\STATE swap them: $\Phi \leftarrow \Phi_s^t$
		\UNTIL{$\nexists$ any approved swapping $\Phi_s^t$}            
	\end{algorithmic}  
\end{algorithm}

Our proposed matching algorithm is summarized in Algorithm~\ref{alg:matching}, which consists of a preliminary interaction and a swapping operation. As for which UTs in ${A}_i$ to accept when there are available seats for ${A}_i$, we summarize the rules in Algorithm~\ref{alg:select} briefly.      

When Phase I is over, we get a preliminary matching result $\tilde{\Phi}$. If there is a blocking pair, i.e., a UT and a UR prefer each other to their current partner, then the UT is free to move to the UR and the UR is free to kick out another UT (if necessary) to make space for the UT. Considering that we do not allow any UT or UR to go outside the system, nor can any UT remain unmatched, we sort to a weaker notion of stability, namely \textit{pairwise stability}. 
First, we define a \textit{swap matching}, denoted by $\Phi_s^t=\{\Phi \setminus \{(UT_s,UR_h),(UT_t,UR_g)\} \} \cup \{(UT_s,UR_g),(UT_t,UR_h)\} $, where $\Phi(UT_s)=UR_h,\Phi(UT_t)=UR_g$, in which $UT_s$ and $UT_t$ switch places while other UTs remain unchanged. Note that one of these two UTs involved in this swap can be a ``seat''. Then we give the definition of pairwise stable as follows~\cite{baron2011peer}

\begin{algorithm}[t] \footnotesize
	\caption{Selecting Algorithm (from List ${A}_i$) }
	\label{alg:select} 
	\begin{algorithmic}
		\STATE \textbf{Data:} $h_{ki}$, $P_k$, $W$, $\sigma$, $e_i$, ${A}_i$\\
		\STATE \textbf{Result:} $\mathcal{A}_i$
		\STATE 1.\textbf{Initialization:} $\mathcal{A}_i=\emptyset$
		\STATE Sort UTs in ${A}_i$ in descending order according to the RP value in $\mathcal{PR}_i$:\\
		$\mathcal{L}_i=\{1th.UT, 2th.UT, \cdots, |{A}_i| th.UT\}$;
		\STATE 2.\textbf{Selection in order:}
		\FOR{$nth.UT=UT_k\in \mathcal{L}_i, n=1:min(e_i,|{A}_i|)$}
		\IF{$\tilde{U}_{i}^{k}>=U_i^{current}$}
		\STATE Add $UT_k$ into $\mathcal{A}_i$.
		\ELSE  \STATE Break;
		\ENDIF
		\ENDFOR
		\STATE $\mathcal{A}_i$ is the list of applicants in ${A}_i$ to be accepted.           
	\end{algorithmic}  
\end{algorithm}

\begin{definition}
A matching $\Phi$ is pairwise stable (PS) if and only if there exists no pair of UTs $(UT_s,UT_t)$, $\Phi(UT_s)=UR_h, \Phi(UT_t)=UR_g$ such that
\end{definition}

\begin{enumerate}
	\item $ \forall m\in \{UT_s,UT_t,UR_h,UR_g\}, U_m(\Phi_s^t) \geq U_m(\Phi)$ and
	\item $\exists m\in \{UT_s,UT_t,UR_h,UR_g\}, U_m(\Phi_s^t) > U_m(\Phi).$
\end{enumerate}

In fact, a swap operation can only occur when all agents of $\{UT_s,UT_t,UR_h,UR_g\}$ ``approve'' it, i.e., the swap must strictly increase at least one agent's utility without decreasing the benefits of all others. In Phase II, we repeat searching for approved swapping pairs and swap the UTs involved until there no longer exist such pairs.  

The convergence of Phase I is ensured by our proposed three constraints in the matching process, because that as long as the total number of ``seats'' $\sum_{i=1}^{M} Q_i >= N$, all UTs can get matched with a UR finally even if the preliminary matching $\tilde{\Phi}$ may be unstable. And the convergence of Phase II is guaranteed due to that the number of ``approved'' swaps is finite. Thus, matching $\Phi$ will converge to a pairwise stable matching in which no agent has the incentive to swap from its current matching partner to another.

\section{An Overlapping Coalition Formation Game for Cooperative Transmission}\label{sec:Cooperative Transmission}

\subsection{Overlapping Coalition Formation Game Formulation}
After the UT-UR association is completed, all UTs have determined their intended data receivers. In the second stage, each UT needs to carefully select its relay UT according to the potential benefits and the required costs. $UT_k$ and all its relay UTs will form a group to perform cooperative beamforming towards the intended UR in the time slot $TS_k$. Due to that each UT can act as the relay of multiple UTs in different time slots, and there is not any utility transfer between UTs, we use a nontransferable utility (NTU) OCF game to figure out an effective overlapping coalition structure, in which each UT can participate in multiple coalitions to achieve higher utility. To better understand the proposed OCF algorithm, we introduce some related definitions in OCF game~\cite{Zhang2017overlapping}.

\begin{definition}
A NTU overlapping coalition formation game is defined as $G=(\mathcal{N},v)$, where $\mathcal{N}$ is the set of players in this game, and $v$ is the utility function. Note that $v(\emptyset)=0$.
\end{definition}

\begin{definition}
An overlapping coalition structure $\Pi$ over $(\mathcal{N},v)$ is defined as a set list: $\Pi=\{C_{1},C_{2},\cdots, C_{u}\}$, where $u$ is the number of coalitions, $\forall 1\leq x \leq u, C_x \subseteq \mathcal{N}$, $\cup_{x=1}^{u} C_x=\mathcal{N}$, and as the coalitions can be overlapped, $\exists x\ne x^{'}, C_x \bigcap C_{x^{'}} \ne \emptyset$.
\end{definition}

\begin{figure}[t]
	\centering
	\includegraphics[width=6.2cm,height=5cm]{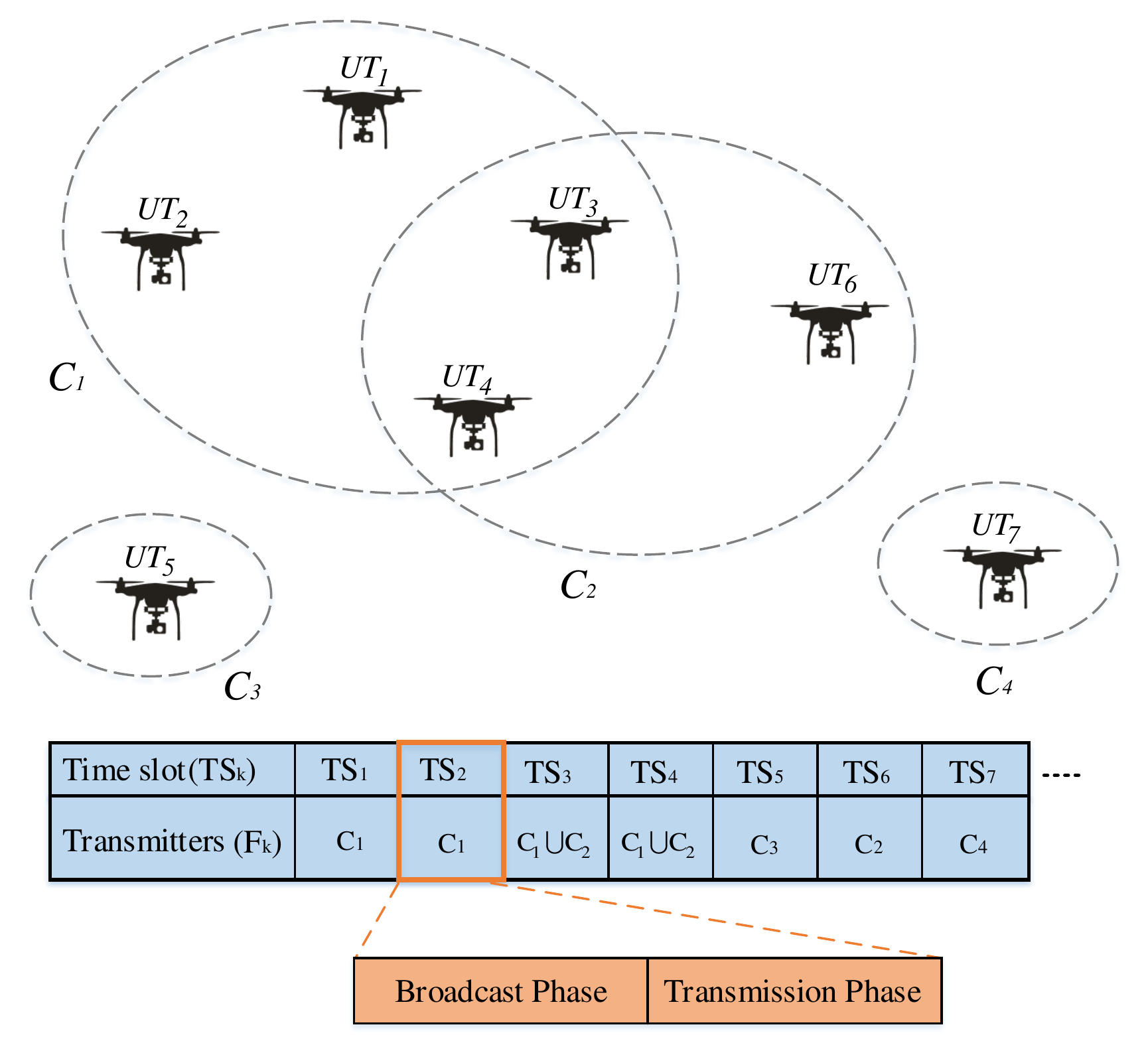}\vspace{-0.3cm}
	\caption{Illustration of an overlapping coalition structure.} 
	\label{fig:coalition structure}\vspace{-0.4cm}
\end{figure}

Corresponding to our investigated scenario, the OCF game players are $N$ UTs. For each $UT_k$, it may have some allies who can act as its relays during time slot $TS_k$, and we denote the group of all its allies plus itself by a set $F_k$. In time slot $TS_k$, all members in $F_k$ will perform the cooperative beamforming together. Given an overlapping coalition structure, each UT can assist the data transmission of some specific UTs in the corresponding time slots. For example, as illustrated in Fig.~\ref{fig:coalition structure}, there are seven UT players and they form a four-coalition overlapping coalition structure. $UT_1$ and $UT_2$ have a same group, i.e., $F_1=F_2=C_1=\{UT_1,UT_2,UT_3,UT_4\}$. $UT_3$ and $UT_4$ have a same group, i.e., $F_3=F_4=C_1\cup C_2=\{UT_1,UT_2,UT_3,UT_4,UT_6\}$, and similarly, there are $F_5=C_3=\{UT_5\}$, $F_6=C_2=\{UT_3,UT_4,UT_6\}$, $F_7=C_4=\{UT_7\}$.

Next, we analyze the payoff and costs of UTs during the cooperative beamforming transmission. For $UT_k$ matched with $UR_i$, in its broadcast phase, there exist a secrecy rate loss due to the fact that $UE_s$ may overhear the information transmission. We define the cost function as the maximum loss value among all UEs, i.e.,
\begin{align}
c_{k} = \frac{1}{2}\max_{s\in\mathcal{S}} \log (1+\frac{P_b |h_{ks}|^2}{\sigma_s^2}),
\end{align}
where $1/2$ represents that this phase occupies half of the time slot. Then in the transmission phase, when $UT_k$ and its allies send a weighted version of the message with the optimal weight vector $\boldsymbol{w}^{*}$, the achievable secrecy rate $C_{k}$, i.e., the payoff of $UT_k$, is given by

\begin{equation}
C_{k}=\frac{1}{2} \log (\alpha +\frac{ | \left( \boldsymbol{w}^{*}\right)^{\dagger} \boldsymbol{h}_{TR_i}|^2}{\sigma_i^2}),
\end{equation}      
where $\alpha=1+{P_b |h_{ki}|^2}/{\sigma_i^2}$. Note that ${P_b |h_{ki}|^2}/{\sigma_i^2}$ is the received SNR at the destination $UR_i$ in the broadcast phase. 

Furthermore, we suppose there exists a minimum SNR threshold above which the relays can effectively decode signals, denoted by $\hat{\gamma}$. Thus, given a coalition structure, in order that all the members of $F_k$ can successfully decode the signal of $UT_k$ and the secrecy loss in this phase can be minimized, the broadcast power $P_b$ should be $P_b={ \hat{\gamma} \,\sigma^2 }/{ |h_{k \tilde{k}}|^2 }$, where $h_{k \tilde{k}}$ is the channel gain between $UT_k$ and $UT_{\tilde{k}}$ who is the furthest ally of $UT_k$ in set $F_k$. If the required $P_b$ exceeds the transmit power budget $P_0$, this $F_k$ can not perform cooperative beamforming successfully and we define the utility in such case as minus infinity. Moreover, in another case where $1<|F_k|< S+1$, the utility is also minus infinity since in this case no optimal weight vector $\boldsymbol{w}^{*}$ can be found in~(\ref{equation7}). Therefore, when a UT is very far from all other UTs, or its neighbors are not enough to form a coalition, it probably chooses to transmit alone, in which case the utility expression is the same as the preference function of $UT_k$ in the matching stage.

In a nutshell, given an overlapping coalition structure, we define the utility function $v$ of $UT_k$ as
\begin{align}
&v_k=
\begin{cases}
\left[ C_k-c_k\right]^{+}, &P_b \leq P_0, \,\,|F_k| \geq (S+1)   \\
U_k^i  ,  &|F_k|=1 \\
-\infty, &otherwise.
\end{cases} 
\label{equation15}  
\end{align}

In addition, we define the total utility $u(\Pi)$ under an overlapping coalition structure $\Pi$ as the sum of all individual utilities, i.e., $u(\Pi)=\sum_{k \in \mathcal{N}} v_k$.

\begin{algorithm}[t] \footnotesize
	\caption{The Overlapping Coalition Formation Algorithm}
	\label{alg:OCF} 
	\begin{algorithmic}
		\STATE \textbf{Initialization:}  $\Pi=\{C_{1},C_{2},\cdots, C_{N}\}, C_m=\emptyset$, for $m=1:N$.
		\STATE 1. For $k=1:N$, identify all potential partners of $UT_k$ and put them into coalition $C_k$.   
		\STATE 2. Correct the coalitions: 
		\STATE If $\exists k, |C_k|<S+1$, then $C_k=\{UT_k\}$, and kick out $UT_k$ from $C_p$(if $UT_k$ exists in $C_p$), $\forall  p\neq k, p\in \mathcal{N}$
		\STATE If $\exists m,n, C_m=C_n, m<n$, remove $C_n$ from structure $\Pi$.
		\STATE A preliminary structure $\Pi=\{C_{1},C_{2},\cdots, C_{u}\}$ is obtained, where $u$ is the number of coalitions after correcting.
		\REPEAT 
		\FOR{$UT_k, k=1:N$}  
		\STATE it randomly select $C_a\in \{C_j|k\in C_j, C_j\in \Pi\}$ and $C_b\in \{C_j|k\notin C_j, C_j\in \Pi\}\cup \emptyset$.
		\STATE $\Pi_{Quit}\triangleq \{\Pi \setminus C_a\}\cup \{C_a \setminus\{i\}\}$
		\IF{$v_k(\Pi_{Quit})\geq v_k(\Pi)$and $u(\Pi_{Quit})\geq u(\Pi)$}
		\STATE $\Pi_k=\Pi_{Quit}$
		\ELSE 
		\STATE $\Pi_{Join}\triangleq \{\Pi \setminus C_b\}\cup \{C_b \cup\{i\}\}$
		\IF{$v_k(\Pi_{Join})> v_k(\Pi)$and $u(\Pi_{Join})\geq u(\Pi)$} 
		\STATE $\Pi_k=\Pi_{Join}$ 
		\ELSE 
		\STATE $\Pi_{Switch}\triangleq \{\Pi \setminus \{C_a, C_b\}\}\cup \{C_b \cup\{i\}\}\cup \{C_a\setminus\{i\}\}$
		\IF{$v_k(\Pi_{Switch})> v_k(\Pi)$and $u(\Pi_{Switch})\geq u(\Pi)$} 
		\STATE $\Pi_k=\Pi_{Switch}$ 
		\ELSE
		\STATE $\Pi_k=\Pi$
		\ENDIF
		\ENDIF
		\ENDIF	             
		\ENDFOR
		\STATE Select a structure with the highest total utility from a set $\Omega\triangleq \{\Pi_1, \Pi_2,\cdots,\Pi_N\}$ and set it as the new structure:\\
		$\Pi=arg \max_{\Pi_k\in\Omega} \{u(\Pi_k)\}$	     
		\UNTIL{$\forall k\in \mathcal{N}, \Pi_k=\Pi$}          
	\end{algorithmic}  
\end{algorithm}
 
\subsection{Solution for the Overlapping Coalition Formation Game}
To update the overlapping coalition structure, we define three basic operations for a UT, which are \textit{Join, Quit, Switch}. \textit{Join} is joining a coalition it doesn't belong to. \textit{Quit} is quitting from a coalition it belongs to. \textit{Switch} is switching from a current coalition to another new one. Since a UT's join or leave may influence some related UTs' utilities, we take both individual and the total utility into account when we conduct these operations in our algorithm. Before describing the proposed OCF algorithm, the concept of \textit{stability} in an OCF game is introduced.

\begin{definition}
An overlapping coalition structure is stable if for $\forall k \in \mathcal{N}$, $UT_k$ can not make any feasible operations, including \textit{Quit}, \textit{Join} or \textit{Switch} move.
\end{definition} 

To achieve a stable overlapping coalition structure, we proposed the OCF algorithm, which is summarized in Algorithm~\ref{alg:OCF}. After a specially designed initialization, $UT_k$ makes a decision whether to quit from a coalition, or join a coalition, or switch from a coalition to another one, or make no change. The structure with the highest total utility is set as the new coalition structure. All UTs repeat this process until any UT would stay in the current coalitions and make no change, because making any move while others remain the same won't bring any benefits.

Next, we prove that the proposed algorithm can achieve a stable overlapping coalition structure, after a finite number of iterations.
\begin{theorem} 
Our proposed OCF algorithm converges to a stable overlapping coalition structure with probability 1.
\end{theorem}

\begin{IEEEproof}
(Convergence) Given the number of players is finite, the total number of possible overlapping coalition structures is finite. In Algorithm~\ref{alg:OCF}, we use a sequence $\{\Pi^{(1)}\to\Pi^{(2)}\to \Pi^{(3)}\to\cdots\}$ describe the evolution of coalition structure. Since each structure with the same highest total utility has an equal chance of being set as the new coalition structure, the evolution process won't keep repeating the previous structure all the time. In addition, every time a UT makes a move (\textit{Quit}/\textit{Join}/\textit{Switch}), a new coalition structure different from the last one will form. So we can describe the evolution of coalition structure such as $\{\Pi_a \to \Pi_b \to \Pi_b \to \Pi_c \to \cdots\}$, where $\Pi_a \neq \Pi_b, \Pi_a \neq \Pi_c, \Pi_b \neq \Pi_c$. Then we explain why the case $\{\Pi_a \to \Pi_b \to \Pi_b \to \Pi_a \to \cdots\}$, in which a "circle" appears, won't happen. Supposing a evolution process with a "circle" just like $\{\Pi_a \to \Pi_b \to \Pi_b \to \Pi_a \to \cdots\}$, we denote the moves from $\Pi_a$ to $\Pi_b$ and from $\Pi_b$ to $\Pi_a$ as $Move_1$ and $Move_2$, respectively. Then all possible combinations of the $(Move_1,Move_2)$ are $(Quit,Join), (Join,Quit), (Switch,Switch)$ (note that these two moves must be of a same UT). If the moves is $(Quit,Join)$, according to the move rules, there must be $u(\Pi_b)\ge u(\Pi_a), v(\Pi_b)\ge v(\Pi_a); u(\Pi_a)\ge u(\Pi_b),v(\Pi_a)> v(\Pi_b)$, which is a contradiction obviously. Similar proof can be made for another two combinations. Therefore, a ``circle'' won't appear in the evolution of coalition structure. In conclusion, during the iteration process, the coalition structure may stay unaltered but it won't turn back among a finite structure set. Therefore, our proposed algorithm will converge to a final coalition structure.

(Stability) Once the algorithm converges to a final structure $\Pi_{final}$, it must be stable, because if $\Pi_{final}$ is unstable and $UT_k$ intends to make a move, there is $\Pi_k\neq \Pi$, which contradicts the fact that the algorithm terminated.
\end{IEEEproof}    

\section{Performance Evaluation}\label{sec:performance evaluation}
\subsection{Simulation Setup}
\begin{figure}[t]	
	\begin{minipage}[t]{0.45\linewidth}
		\centering
		\includegraphics[width=4cm,height=2.6cm]{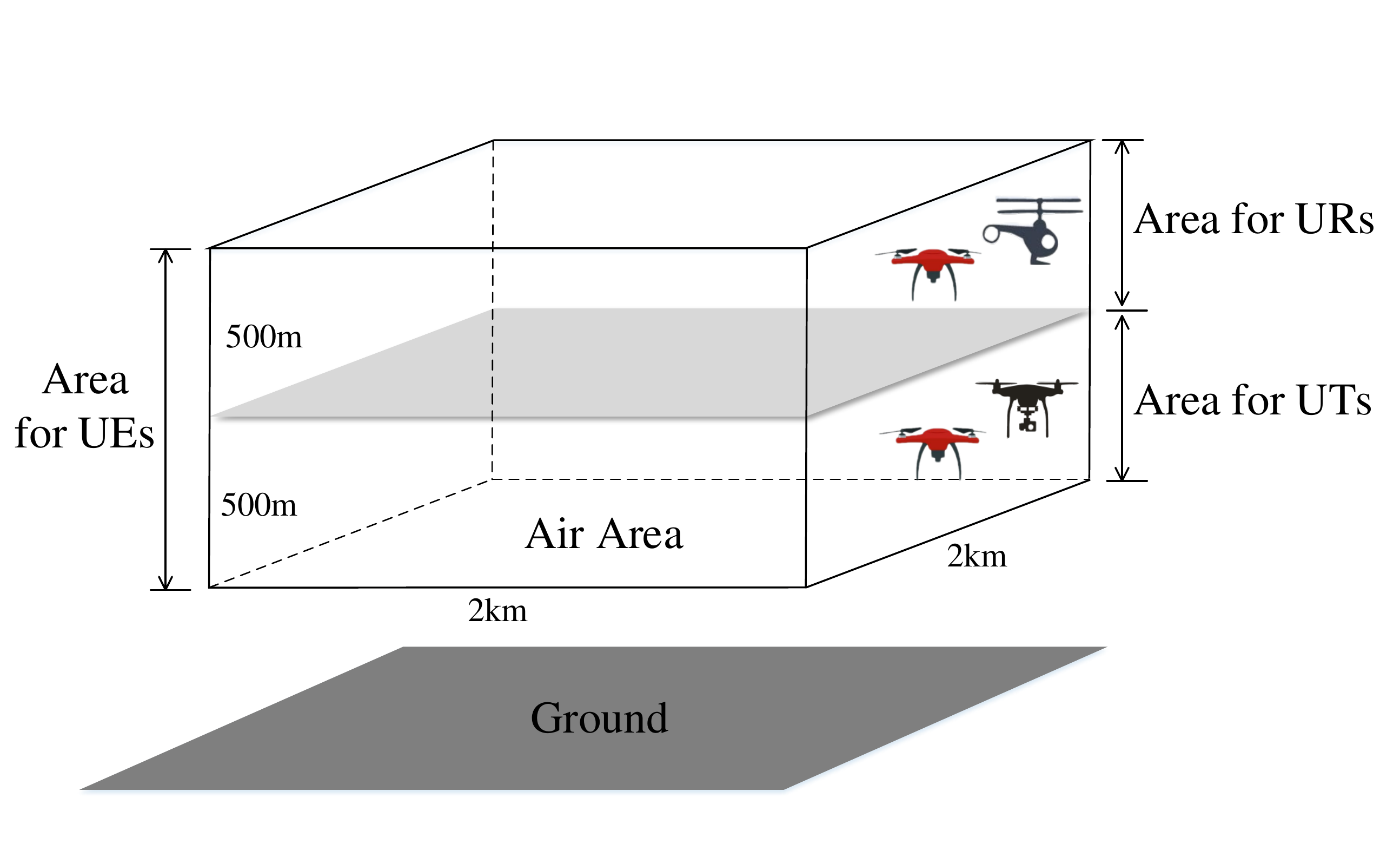}
		\caption{Illustration for the distribution settings of UTs, URs and UEs.}
		\label{fig:deployment}
	\end{minipage}
	\begin{minipage}[t]{0.45\linewidth}
		\centering
		\includegraphics[width=4cm,height=2.8cm]{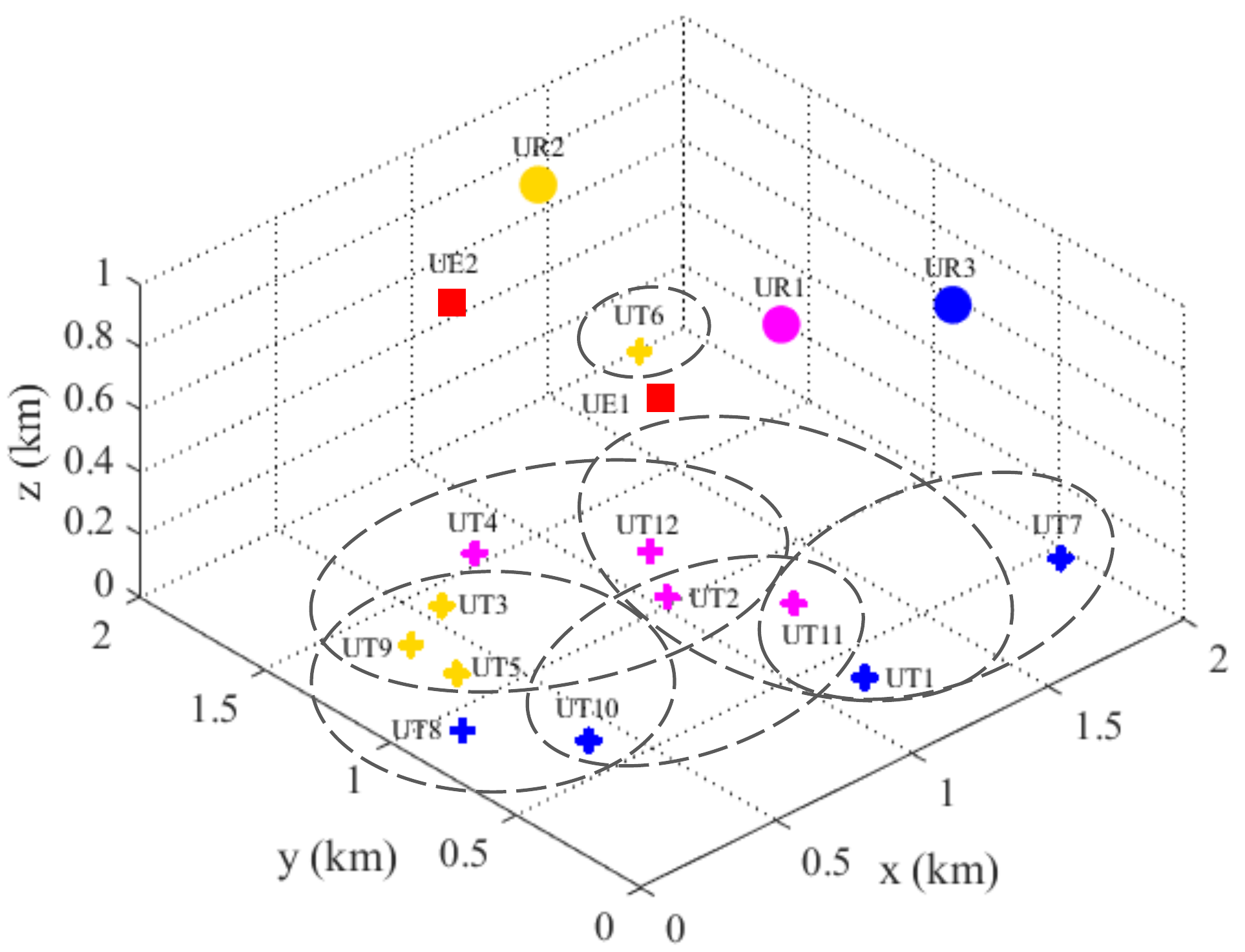}
		\caption{A snapshot of the final structure with N=12 UTs, M=3 URs, R=2 UEs and Q=4.}
		\label{fig:snapshot}
	\end{minipage}
\end{figure}

As illustrated in Fig.~\ref{fig:deployment}, the UTs are randomly distributed in a rectangular area of $2km\times 2km\times 500m$ in the air, the URs are randomly distributed in a higher area of $2km\times 2km\times 500m$, and the UEs are randomly located in the $2km\times 2km\times 1km$ area which involves the areas of UTs and URs. The simulation parameters are given in Table~\ref{table:parameters} unless otherwise specified. The URs and UEs are assumed to have the same noise power, and all URs are assumed to have the same quota. In order to obtain more reliable simulation results, for a fixed number of UTs, URs and UEs, we repeat generating random location layouts by 100 times and average the simulation results.

Fig.~\ref{fig:snapshot} shows a snapshot of a pairwise-stable matching and overlapping coalition structure resulting from our proposed algorithms. The UTs, URs, and UEs are represented by crosses, circles, and red squares respectively. We use the same color denote the matching relationship, i.e., the UTs are matched with a UR with the same color. The gray dotted circles represent the overlapping coalition structure that UTs eventually formed. It shows that the UTs form a complex overlapping coalition structure under this position layout. Some UTs such as $UT_2$ and $UT_{11}$ participate in multiple coalitions, which means they own lots of allies when performing cooperative beamforming, of course, they also need to assist their allies in turn during these allies' time slots. Therefore, it's a mutually beneficial way to work in coalitions. In addition, there may exist UTs that prefer to work alone, just like $UT_{6}$, this is because it is too far away from other UTs to benefit from the cooperation.

\begin{table}[t] \footnotesize
	\centering  
	\caption{Simulation parameters.}
	\label{table:parameters}
	\begin{tabular}{ccc}  
		\hline
		Parameter &Symbol &Value \\ 
		\hline 
		Number of UTs & $N$ & 12\\
		Number of URs & $M$ & 3\\
		Number of UEs & $R$ &2\\
		Quota of URs & $Q$ &4\\
		System bandwidth &$W$ &100kHz \\        
		Noise power&$\sigma^2$ &-60dBm\\       
		UT transmission power budget &$P_0$ &10dBm \\ 
		Path loss exponent&$\alpha$ &2\\
		SNR threshold & $\hat{\gamma}$ & 10dB\\
		\hline
	\end{tabular}
\end{table}

\subsection{Performance Analysis}
To evaluate the effectiveness of our proposed matching algorithm and OCF algorithm, which are labeled as PMA and OCFA respectively, we compare them with other practical schemes. First, we fix the matching scheme in stage 1 to be PMA, and compare the OCFA with the other three practical schemes under various system parameters in terms of the total utility of all UTs, which represents the secrecy performance of the whole network when transmitting data. Then, we focus on validating the performance of PMA compared to other typical matching schemes in terms of the social welfare of all UTs and URs. In addition, we give some necessary analyses for all simulation results.	

Now we compare the proposed OCFA with the following three transmission schemes, note that the same transmit power budget $P_0$ is applied to the individual or the whole coalition:

a) \textbf{Alone Scheme} (AS), where each UT transmits its data by itself without any other cooperating relay in the cooperative transmission stage, which is a non-cooperative approach.

b) \textbf{Full Group Scheme} (FGS), where each UT transmits its data with all the UTs within its effective communication circle being its relays in the cooperative transmission stage.

c) \textbf{Disjoint Coalition Scheme} (DCS), where the UTs form disjoint coalitions in the cooperative transmission stage. We employ the $\textit{q-merge and 2-split}$ scheme proposed in~\cite{mochaourab2017distributed}. The parameter $q$ is the maximum number of coalitions that merge into a larger coalition and is set from 2 to 6. We choose the maximum performance value in this scheme for comparison in our evaluation.

\begin{figure*}[t]
	\centering
	\begin{minipage}[t]{0.24\linewidth}
		\centering
		\includegraphics[width=4cm,height=3cm]{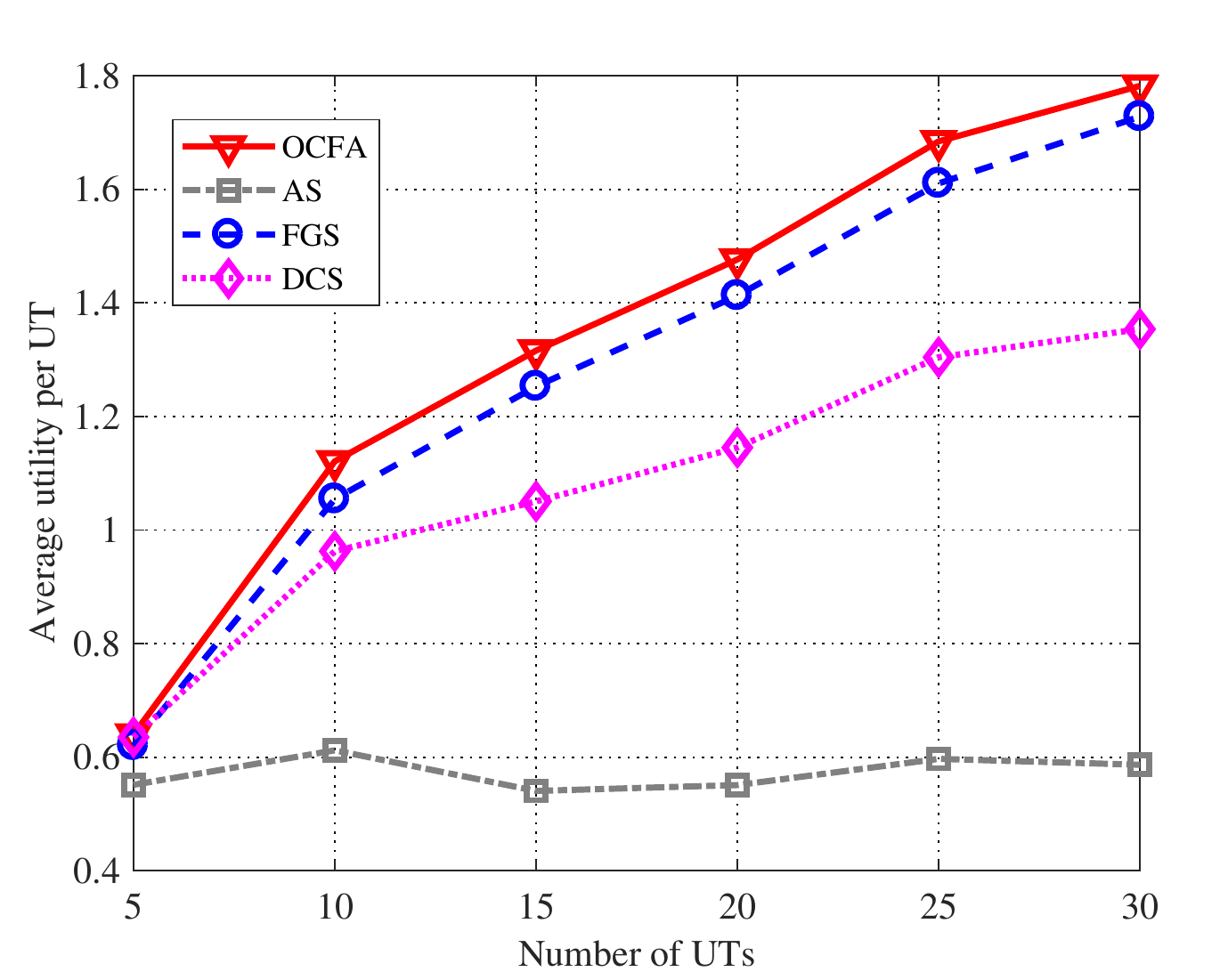}
		\caption{Average user utility comparison of four transmission schemes with different number of UTs.}
		\label{fig:N_TotalUtility_sub2}
	\end{minipage}
	\begin{minipage}[t]{0.24\linewidth}
		\centering
		\includegraphics[width=4cm,height=3cm]{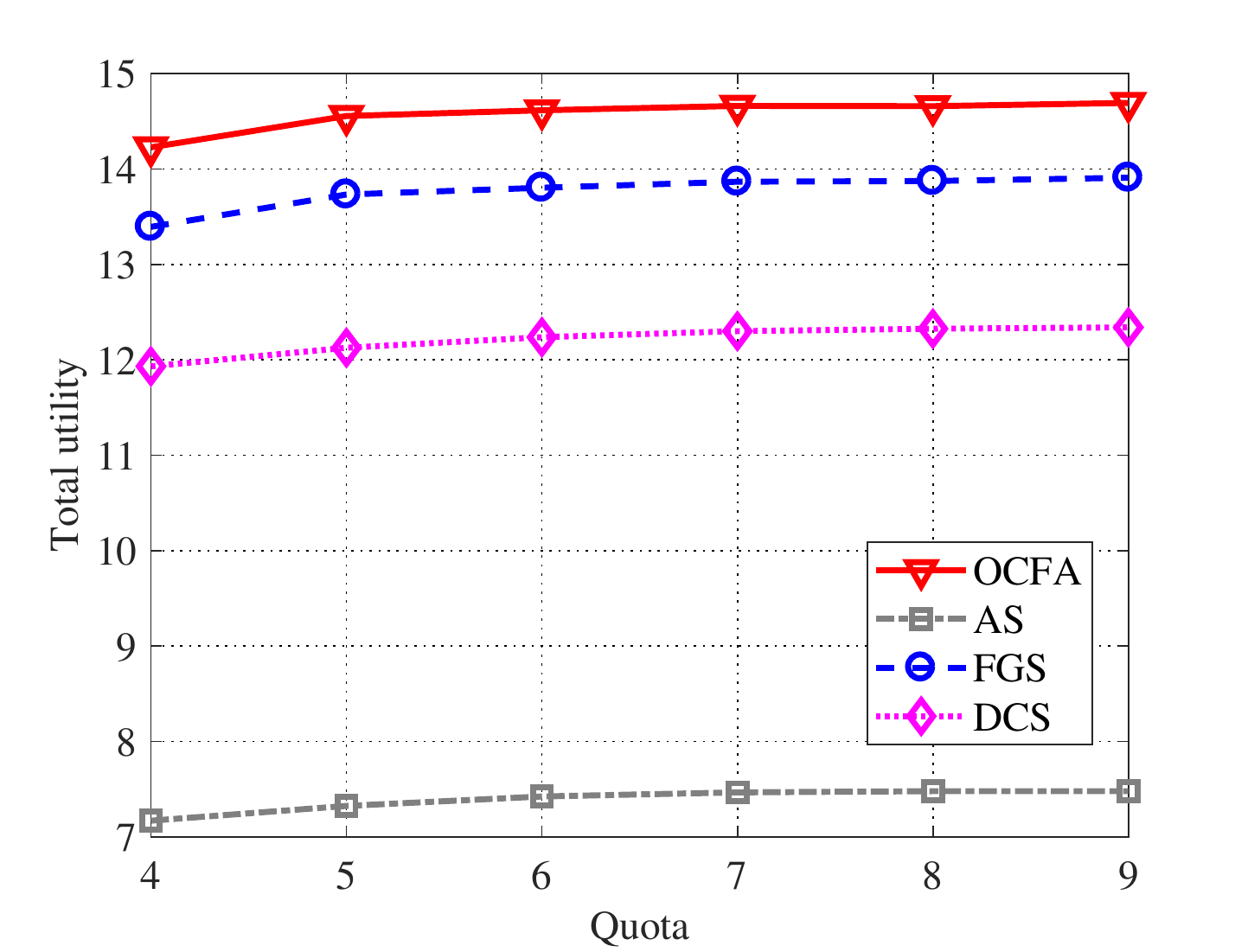}
		\caption{Total utility comparison with different number of $Q$.}
		\label{fig:Q_TotalUtility}
	\end{minipage}
	\begin{minipage}[t]{0.24\linewidth}
		\centering
		\includegraphics[width=4cm,height=3cm]{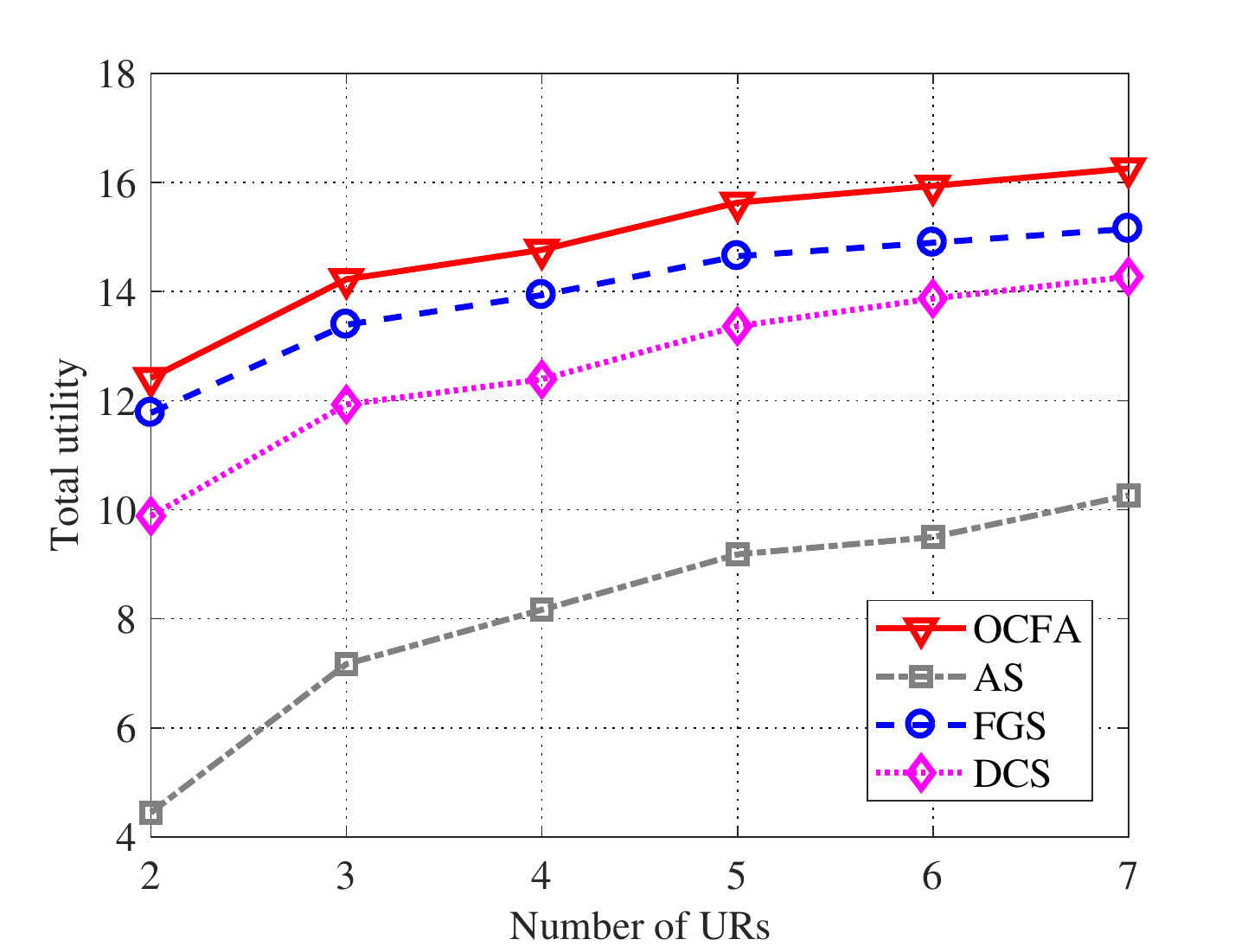}
		\caption{Total utility comparison with different number of $M$.}
		\label{fig:M_TotalUtility}
	\end{minipage}
	\begin{minipage}[t]{0.24\linewidth}
		\centering
		\includegraphics[width=4cm,height=3cm]{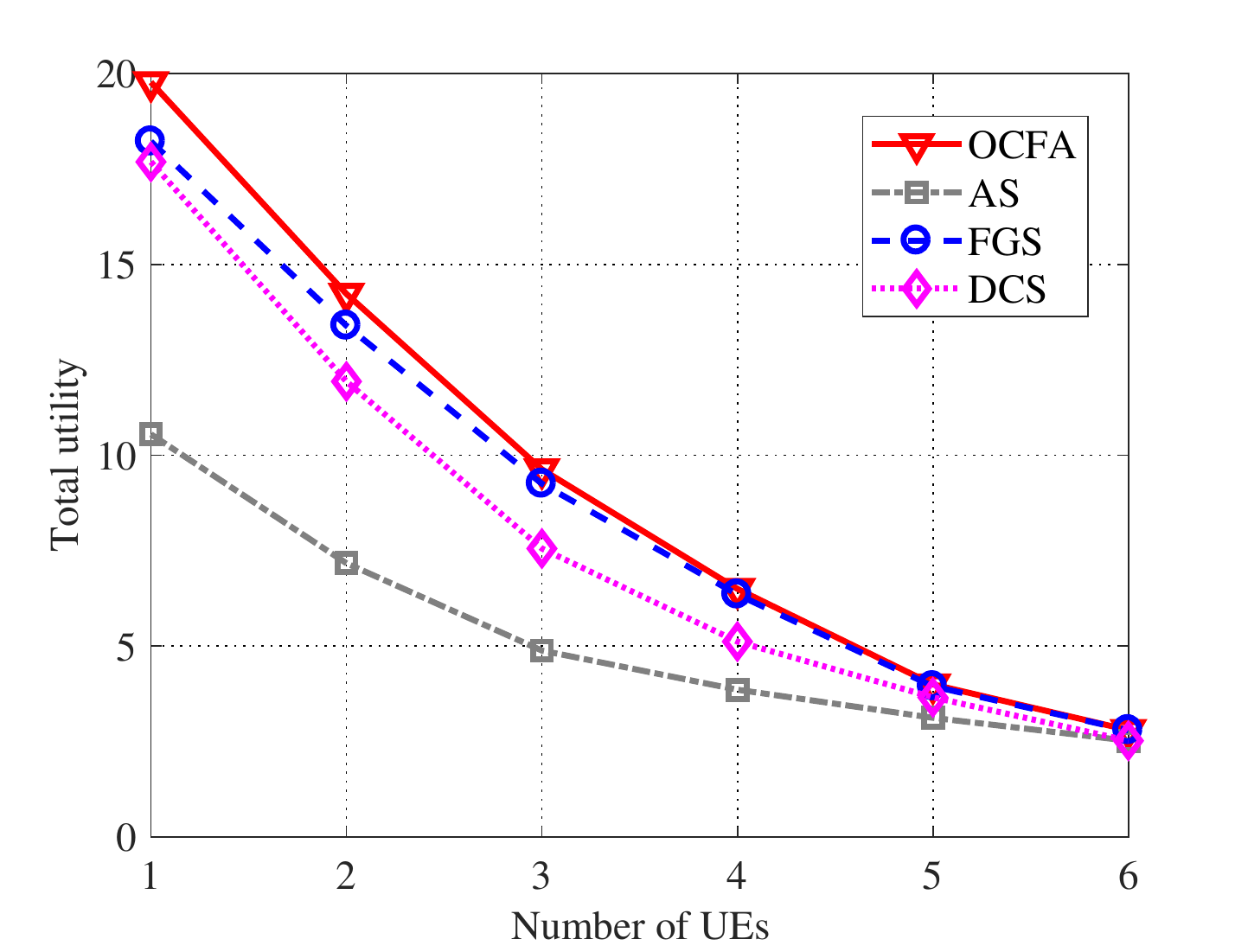}
		\caption{Total utility comparison with different number of R.}
		\label{fig:R_TotalUtility}
	\end{minipage}
\end{figure*}

In Fig.~\ref{fig:N_TotalUtility_sub2}, we plot the average utility per UT of four schemes under varying $N$. When $N$ is from 10 to 30, we can see the average utility per UT of OCFA, FGS, and DCS increase as $N$ becomes larger, while that of AS is almost maintained at a relatively low level. It is because that for the cooperative schemes including OCFA, FGS, and DCS, there are more potential allies within each UT's communication range as the number of UTs increases, which probably results in bigger coalitions and more complex overlapping coalition structure. However, for the AS, in which each UT transmits its data alone during its whole time slot, the increase of neighbors of a UT will not benefit it and may even bring in a competition for the same desired UR in stage 1. Therefore, the cooperative approaches including OCFA, FGS, and DCS are superior to AS, which is a non-cooperative approach. 

\begin{figure*}[t]
	\begin{minipage}[t]{0.24\linewidth}
		\centering
		\includegraphics[width=4cm,height=3cm]{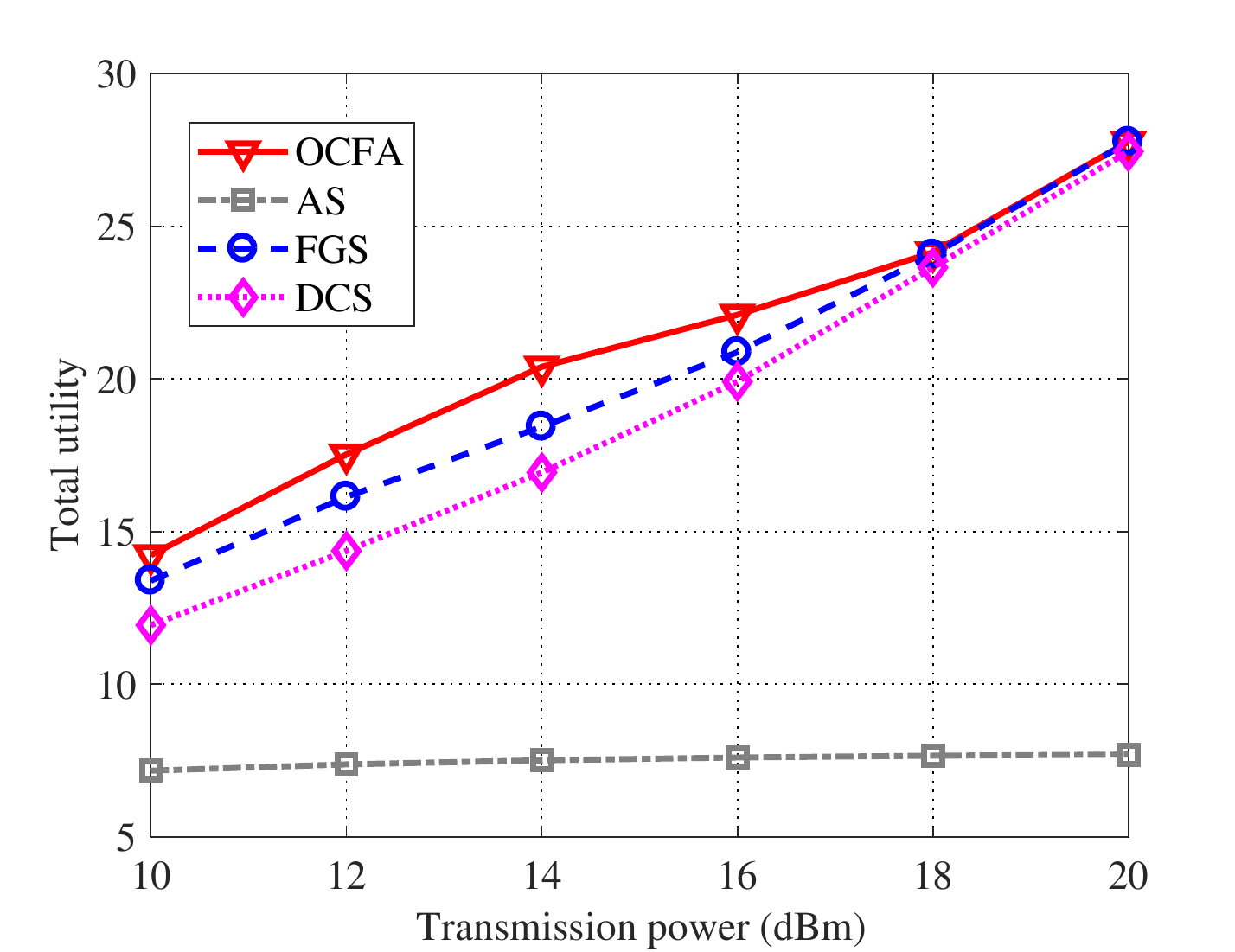}
		\caption{Total utility comparison with different transmission power.}
		\label{fig:P_TotalUtility}
	\end{minipage}
	\begin{minipage}[t]{0.24\linewidth}
		\centering
		\includegraphics[width=4cm,height=3cm]{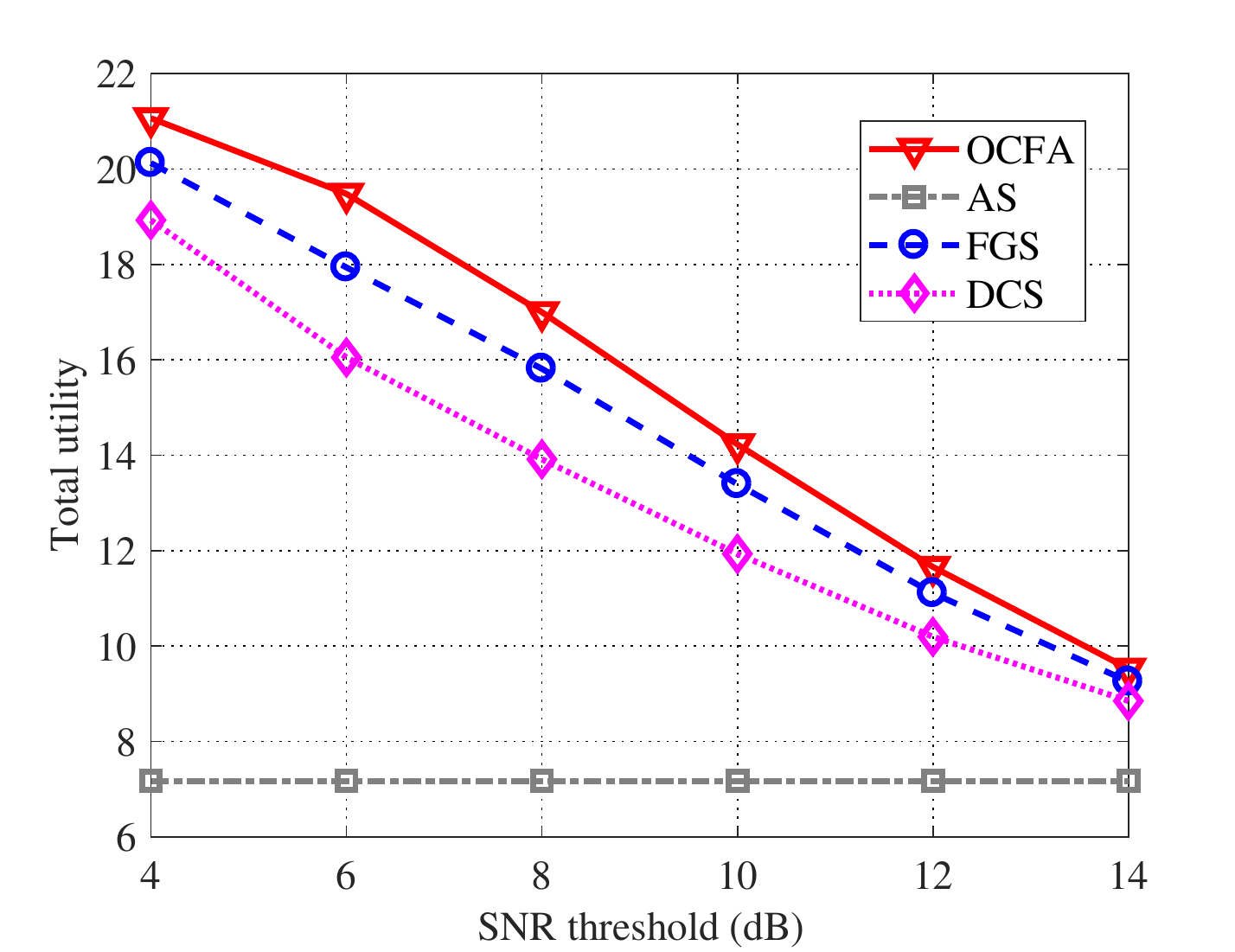}
		\caption{Total utility comparison with different SNR threshold.}
		\label{fig:th_TotalUtility}
	\end{minipage}
	\begin{minipage}[t]{0.24\linewidth}
		\centering
		\includegraphics[width=4cm,height=3cm]{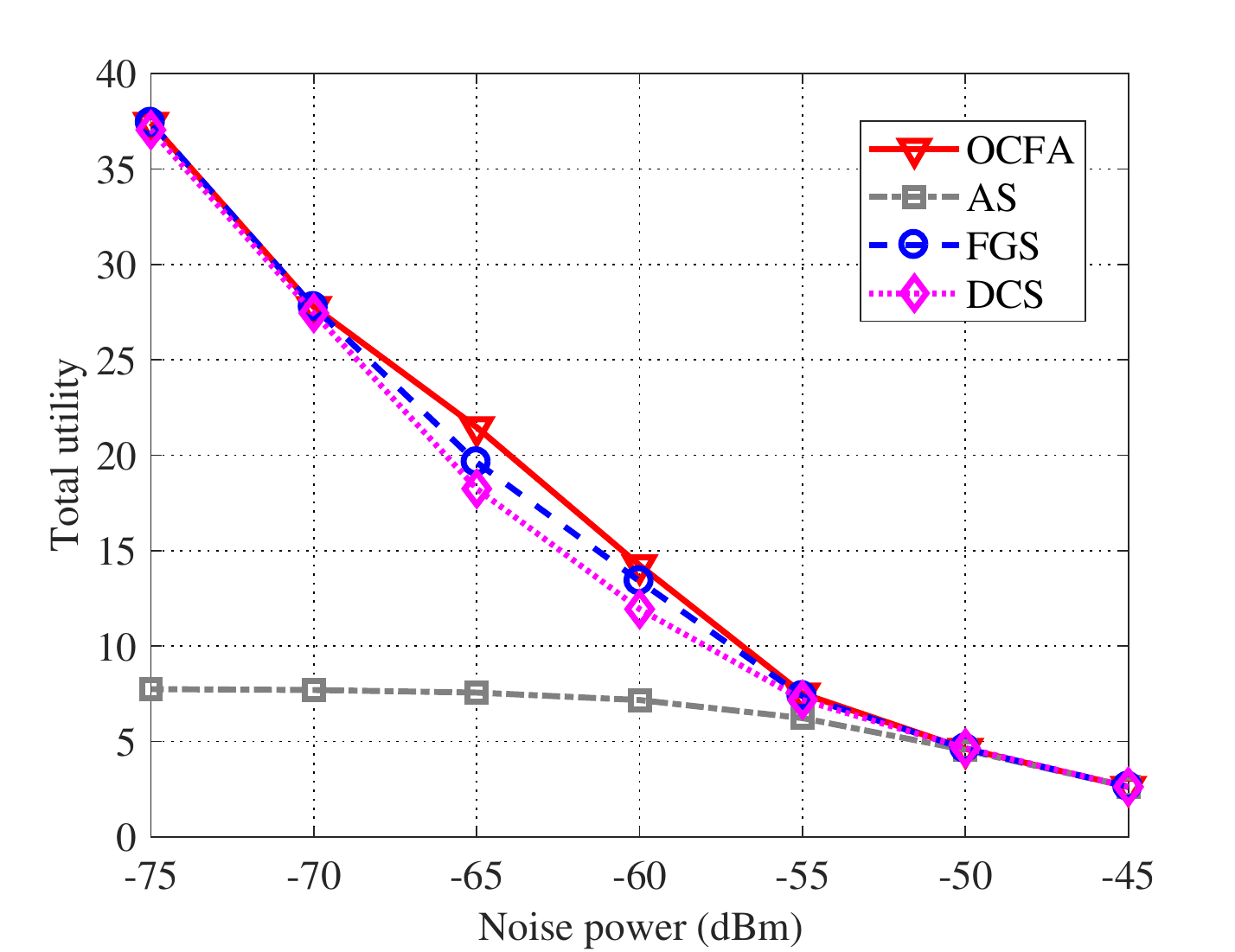}
		\caption{Total utility comparison with different noise power.}
		\label{fig:n2_TotalUtility}
	\end{minipage}
	\begin{minipage}[t]{0.24\linewidth}
		\centering
		\includegraphics[width=4cm,height=3cm]{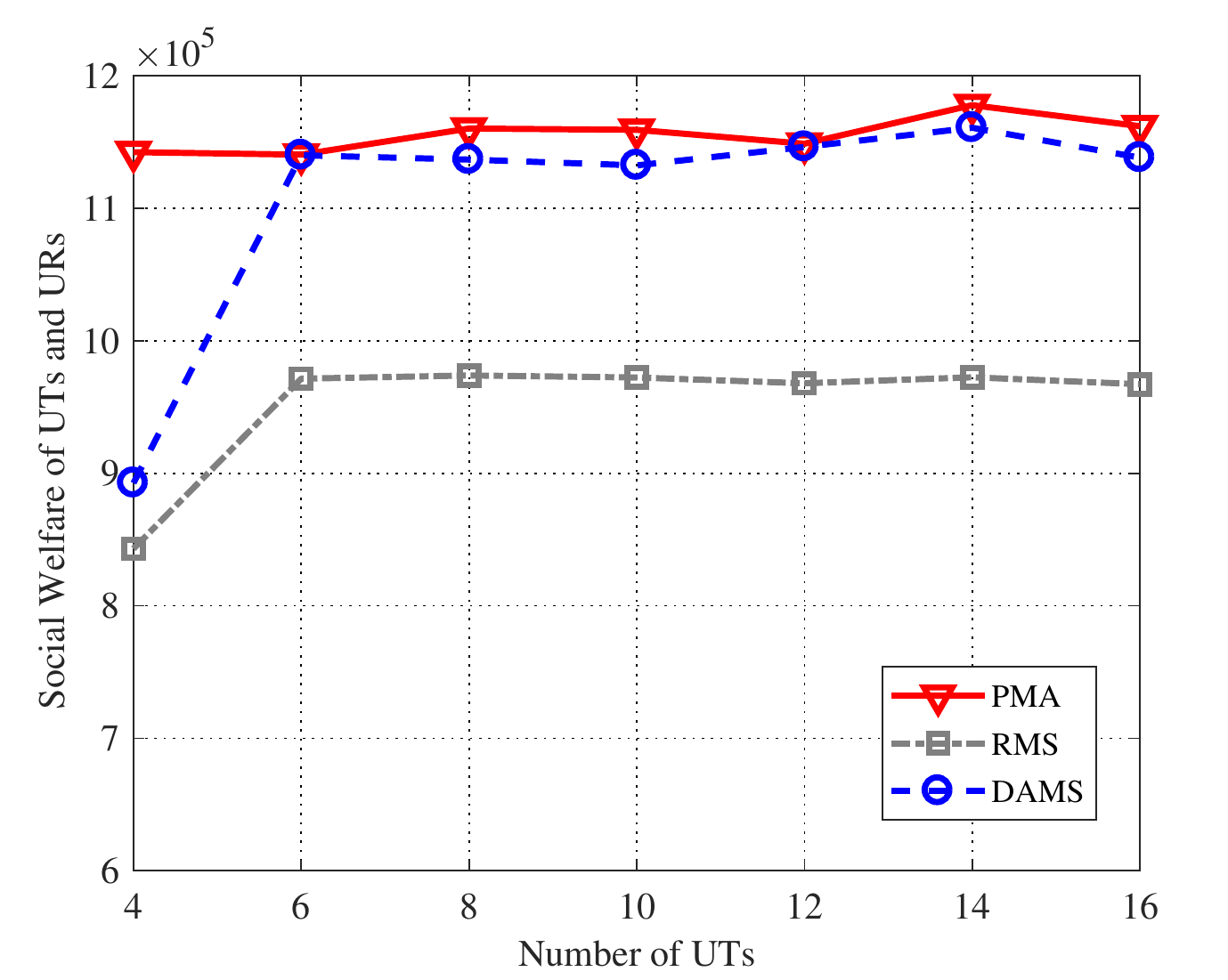}
		\caption{Social welfare comparison of three matching schemes with different number of UTs.} 
		\label{fig:N_SocialWelfare}
	\end{minipage}
\end{figure*}

From Fig.~\ref{fig:Q_TotalUtility}, it is noticed that the total utilities of four schemes have a very slight increase as the quota for each UR increases. It is because that with more provided "seat" in stage 1, a UT has more chance to match the preferred UR, who can establish better channel state with itself in general, resulting in an improvement in the total utility when performing cooperative beamforming in stage 2. However, when a UR can serve more UTs, although there are more ``hole'', a UT can only switch to this ``hole'' when all involved agents approve this operation, which limits big changes in the matching results. Therefore, the impact of quota on the total utility is very slight.   

In Fig.~\ref{fig:M_TotalUtility}, we vary the number of URs to be 2 to 7. As all UTs need to match a UR finally, there must be a constraint of $M\times Q>=N$, thus we adapt $Q$ correspondingly as the number of URs changes ($Q$ is set 6, 4, 3, 3, 3, 2 as the number of URs to be $2\sim 7$). From this figure, we can see OCFA performs better than other schemes with varying $M$. When the number of URs is 7, the total utility of OCFA is larger than that of FGS, DCS and AS about 9\%, 10\% and 54\%, respectively. In addition, the total utilities of four schemes increase as the number of URs increases. This is due to that with more URs in the network, UTs are more likely to match a UR with better channel conditions, thereby improving the utility when performing cooperative beamforming. Besides, the AS still gets the worst performance compared to the other three cooperative approaches. In Fig.~\ref{fig:R_TotalUtility}, we vary the number of UEs from 1 to 6. We can observe that with more UEs distributed in the network, the total utilities of four schemes will decrease, which is in accordance with our forecasts. With more UEs randomly distributed in this area, a UT is more likely to be close to a UE, which leads to a higher secrecy loss for this UT no matter it works alone or in a team, i.e., in a cooperative or non-cooperative way. 

We also investigate the impact of transmit power $P_0$ variation on the total utilities of four schemes. From Fig.~\ref{fig:P_TotalUtility}, we can notice that the total utilities of OCFA, FGS and DCS increase obviously as the transmit power increases, while the AS curve grows very slightly. The reason is that with larger transmit power, UTs work in a cooperative way have a wider communication circle, which means they could own more potential allies to combat against eavesdropping. Despite a part of power loss and secrecy loss resulting from the information broadcast phase, the gains from cooperation outweigh the losses, which leads to an obvious improvement in the total utility. For AS, the increase of transmit power would only result in a slight improvement in the secrecy performance. Furthermore, we can see that when the transmit power is 18dBm, the curves of OCFA and FGS begin to meet. In essence, when the transmit power increases to 18dBm, the effective communication circle radius becomes about 2.5km, which implies a UT can cover nearly all others within its communication range in this network, and all UTs probably form a grand coalition to perform cooperative beamforming. Therefore, when the transmit power is large enough, OCFA and FGS obtain the same total utility in which cases a grand coalition of all UTs will form.    

In Fig.~\ref{fig:th_TotalUtility} and Fig.~\ref{fig:n2_TotalUtility}, we vary the SNR threshold and the noise power to investigate their impacts on the total utility, respectively. As shown in Fig.~\ref{fig:th_TotalUtility}, the total utilities of OCFA, FGS and DCS decrease as the SNR threshold increases, while the total utility of AS has no change. It is easy to understand that with a larger SNR threshold, a UT can not form a coalition with the UTs far away from it because they can not successfully decode and forward the message when performing cooperative beamforming, which leads to a smaller effective communication circle. As for AS, the change of SNR threshold has no effect on it due to its non-cooperative way. Fig.~\ref{fig:n2_TotalUtility} indicates that the total utilities of all four schemes decrease as the noise power increases. With higher noise power and fixed SNR threshold, each UT's communication circle becomes smaller, which means each UT will have fewer potential allies. Then we can notice that when the noise power is about -70dBm or lower, OCFA and FGS show the same performance, in which case a grand coalition of all UTs forms to perform cooperative beamforming. As the noise power becomes higher, all UTs will make their rational decisions to quit from the grand coalition and self-organized into a new complex overlapping coalition structure based on their utilities. When the noise power increases to about -50dBm, as shown in this figure, the cooperative approaches of OCFA, FGS and DCS become no different from AS, in which case each UT forms a single-player coalition due to a much higher broadcast communication cost.

Finally, in order to show the effectiveness of the proposed matching algorithm in the UT-UR association stage, we compare PMA with the classical DA matching algorithm~\cite{gu2015matching}, labeled as DAMS. In addition, a random matching scheme (RMS) is given as a benchmark.  

Fig.~\ref{fig:N_SocialWelfare} indicates that PMA outperforms DAMS slightly in terms of the social welfare of UTs and URs in stage 1. The underlying reason is that in PMA, a UR can choose to reject some first-time applicants who cannot improve its current benefit even if there are left "seats" for them, while in DAMS a UR accepts as many applicants as possible (up to the quota). In our considered network, there may be cases where a UE becomes very close to a UT, and for this UT, which UR to match  makes no difference to its benefit according to the preference function, however, it makes a difference to URs. The UT who does not care which UR to match may be more preferred by another UR rather than the one it is applying to. Thus in PMA, URs can own more options, and the social welfare of UTs and URs can be improved to some extent. 
   
\begin{figure}[t]
	\centering	
	\subfigure[]{
		\begin{minipage}[t]{0.45\linewidth}
			\centering
			\includegraphics[width=4cm,height=3cm]{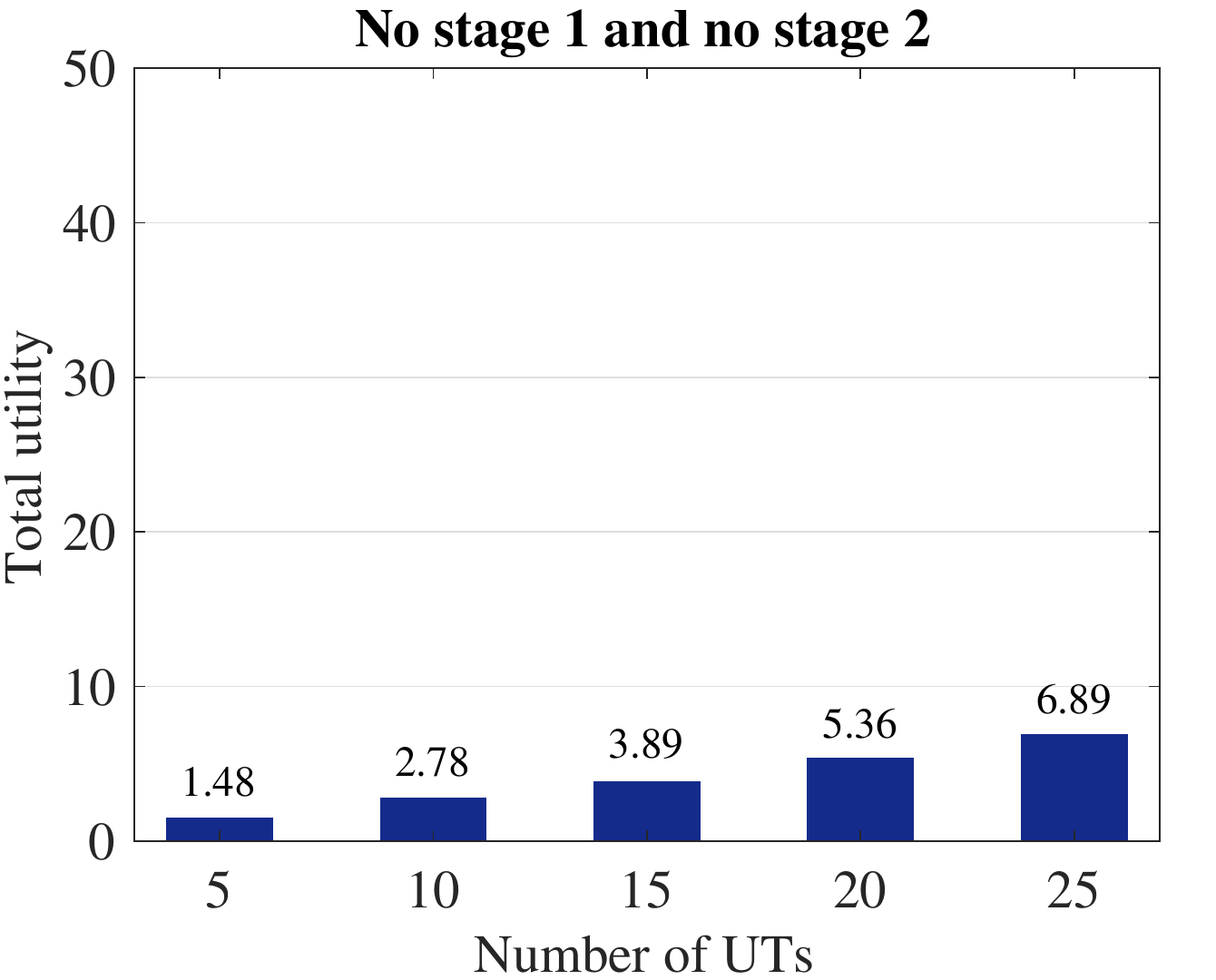}
		\end{minipage}
	}
	\subfigure[]{
		\begin{minipage}[t]{0.45\linewidth}
			\centering
			\includegraphics[width=4cm,height=3cm]{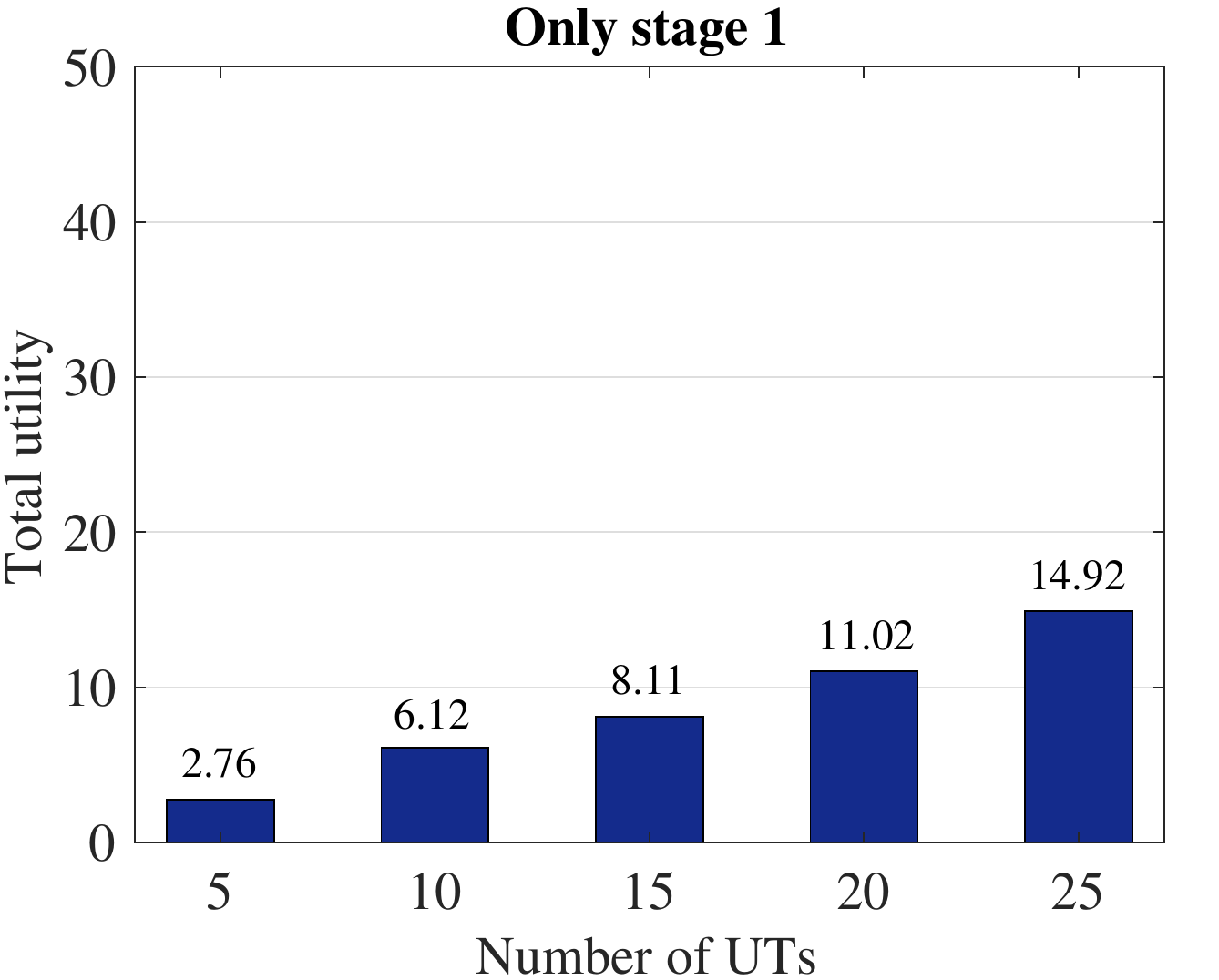}
		\end{minipage}
	} 
	\subfigure[]{
		\begin{minipage}[t]{0.45\linewidth}
			\centering
			\includegraphics[width=4cm,height=3cm]{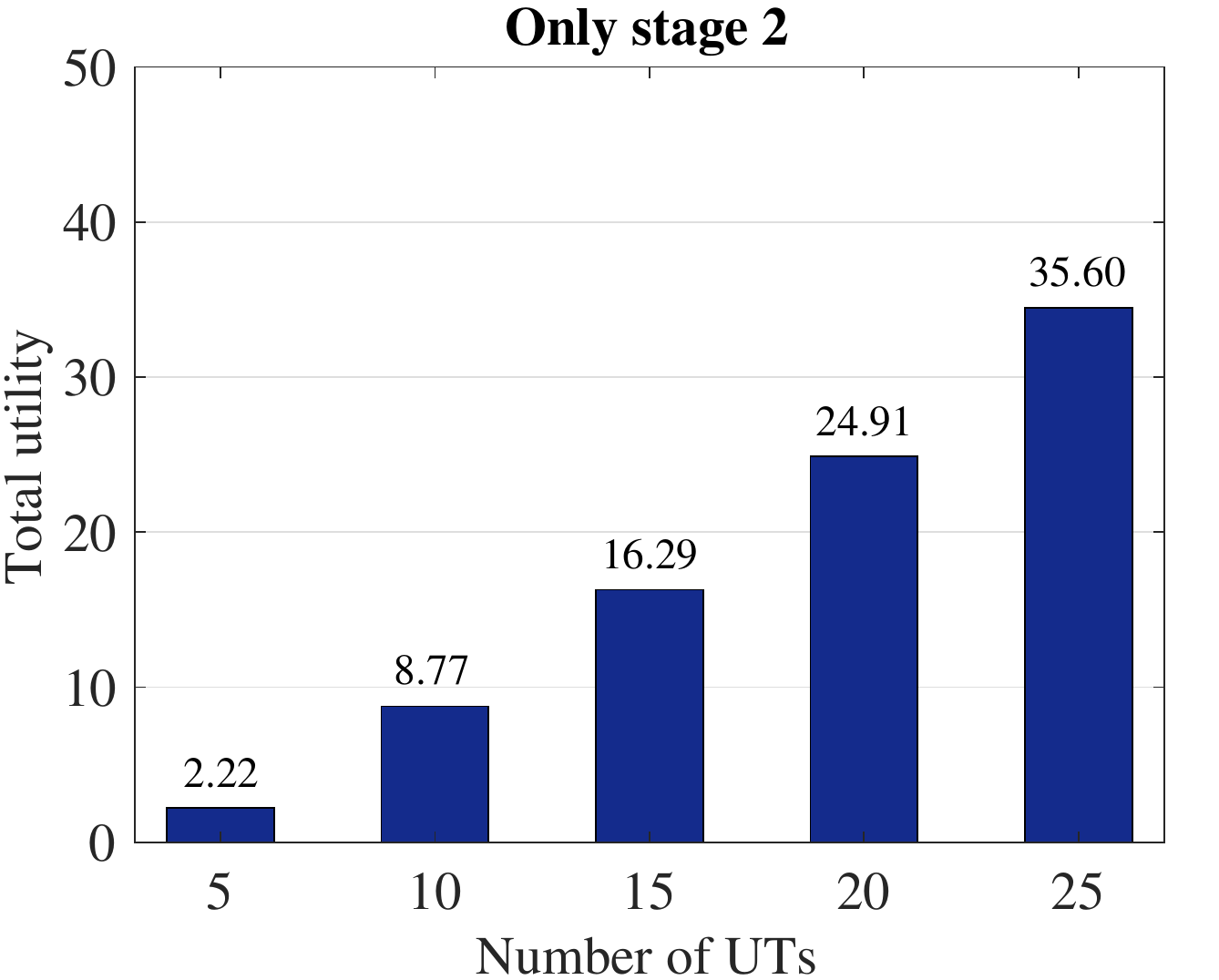}
		\end{minipage}
	}
	\subfigure[]{ 
		\begin{minipage}[t]{0.45\linewidth}
			\centering
			\includegraphics[width=4cm,height=3cm]{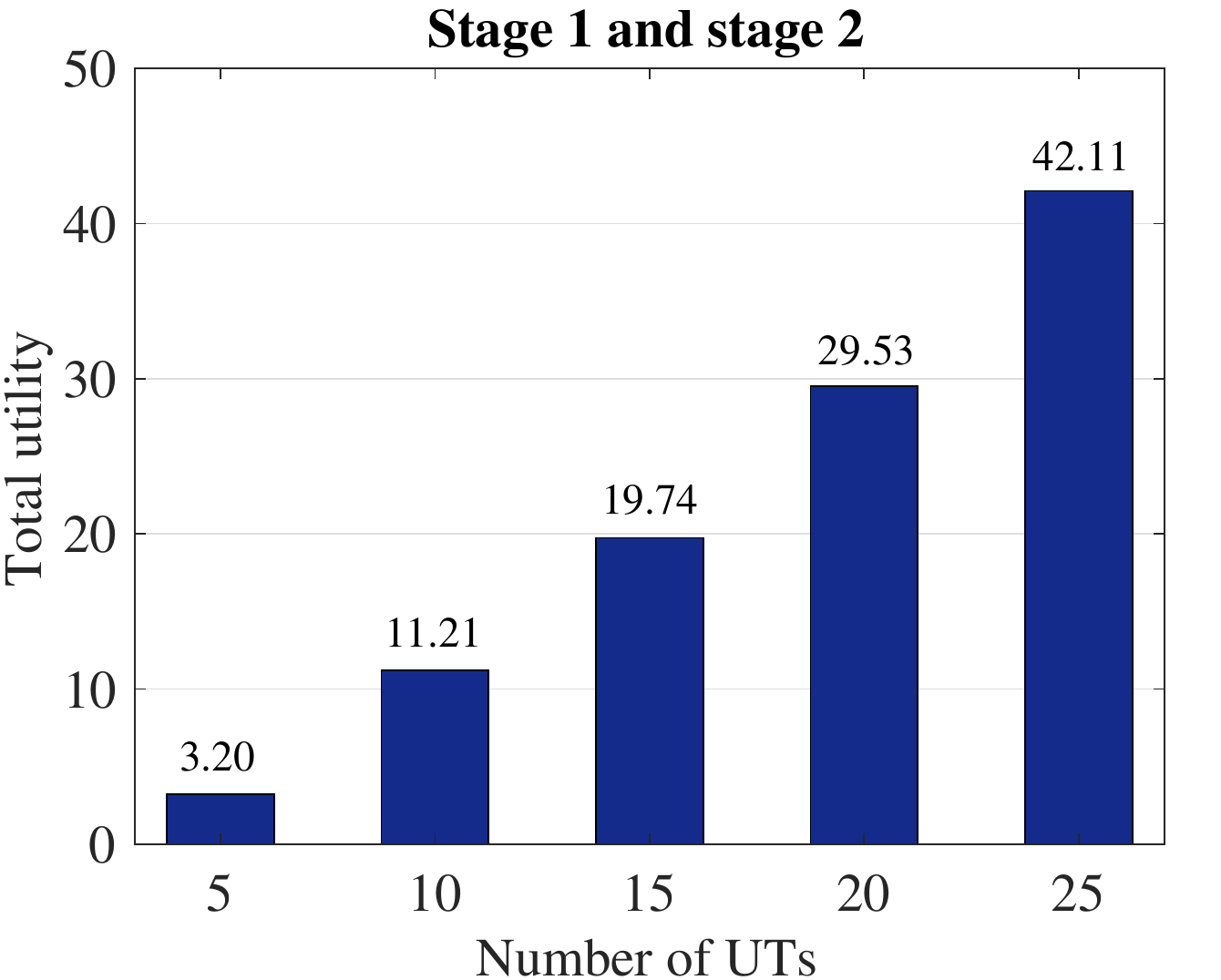}
		\end{minipage}
	}	
	\caption{(a) Total utility when neither stage is employed (random UT-UR association and alone transmission), (b) Total utility when only stage 1 is employed (UT-UR association with the proposed matching game and alone transmission), (c) Total utility when only stage 2 is employed (random UT-UR association and cooperative beamforming with the OCF game) and (d) Total utility when both stage 1 and stage 2 are employed (UT-UR association with the matching game and cooperative beamforming with the OCF game).}
	\label{fig:two_stages}
\end{figure}

In Fig.~\ref{fig:two_stages}, we show the effectiveness of our proposed two-stage framework. As shown in this figure, when neither stage is applied, i.e., random UT-UR association and alone transmission, the total utility is very low. Employing stage 1 can provide a good matching structure for the subsequent transmission to achieve better total utility, and employing stage 2 can further improve the total utility due to the effective overlapping coalition based cooperative beamforming. Furthermore, the combination of these two stages can significantly improve the total utility, thereby greatly enhancing the transmission security in the network.    

\section{Conclusion}\label{sec:conclusion}
In this paper, we investigate the secure transmission problem in a two-tier UAV network. To enhance the PLS of this network, we utilize cooperative beamforming to combat against the UEs and design a two-stage framework consisting of a UT-UR association stage and a cooperative transmission stage. We formulate the UT-UR association problem and the relay selection problem into a many-to-one game and an OCF game, respectively. Then a many-to-one matching algorithm and an OCF algorithm are proposed to solve these two sequential games. The stabilities of these two algorithms are theoretically proved, and extensive simulations are shown to demonstrate the superior performance of our proposed schemes and the effectiveness of our proposed two-stage framework.
\bibliographystyle{IEEEtran}
\bibliography{IEEEabrv,new}
\end{document}